\documentclass[a4paper,11pt]{article}
\usepackage{jheppub} % for details on the use of the package, please see the JINST-author-manual
\usepackage{lineno}
\usepackage{amsmath,amssymb,amsthm}
\usepackage{color}
\usepackage{amsfonts}
\usepackage{tikz-cd}
\usepackage{cancel}
\usepackage{arydshln}
\usepackage{slashed}
\usepackage{feynmp}
\usepackage{xcolor}

\newcommand\xCcancel[2][black]{\renewcommand\CancelColor{\color{#1}}\xcancel{#2}}

\DeclareMathOperator{\tr}{\textrm{tr}}
\DeclareMathOperator{\Tr}{\textrm{Tr}}
\DeclareMathOperator{\Str}{\textrm{Str}}
\DeclareMathOperator{\Ad}{\textrm{Ad}}
\DeclareMathOperator{\ad}{\textrm{ad}}

\arxivnumber{2312.15557} % if you have one

\title{Beta deformed sigma model and strong deformation coupling limit}

\author{Eggon Viana}
\affiliation{Instituto de Fisica Teorica, Universidade Estadual Paulista\\
R. Dr. Bento Teobaldo Ferraz 271, 
Bloco II -- Barra Funda\\
CEP:01140-070 -- Sao Paulo, Brasil\\}

\emailAdd{eggon.viana@unesp.br}

\abstract{We study the beta deformation of the superstring in $AdS_5\times S^5$ at all orders in the deformation parameter, employing the pure spinor formalism. This is necessary in order to study the regime of strong deformation parameter, which in the field side is related to fishnet theories. We compare the pure spinor sigma model approach to the previously known supergravity description. We find a complete agreement. Moreover, the BRST structure of the worldsheet model provides a natural explanation of the peculiar features of the worldsheet model in the fishnet limit. In particular, we study the degeneracy of the sigma model Lagrangian. We show that the BRST structure is responsible for a particularly ``tame" degeneration of the fishnet sigma-model.}

\begin{document}
\maketitle
\flushbottom

\section{Introduction}

The $\mathcal{N}=4$ Supersymmetric Yang-Mills theory has been one of the most studied theories in high energy physics over the last few decades due to the prominent role it plays in the AdS/CFT correspondence \cite{Maldacena:1997}, \cite{Witten:1998}. One important advance in its understanding was the introduction of a marginal deformation by Leight and Strassler, called $\beta$ deformation \cite{strassler}, which preserves $\mathcal{N}=1$ supersymmetry. A generalization to 3-parameters deformation is also known \cite{Fokken_2014} and does not preserves any supersymmetry in general. This deformation has been recently applied to obtain fishnet theories in the field side \cite{Gurdogan:2015csr}, which arise in the limit of strong deformation.

The gravity dual of such deformed theories was found by Maldacena, Lunin and Frolov \cite{Lunin_2005},\cite{Frolov_2005}. Worldsheet description in the pure spinor formalism was proposed in \cite{andrei}. However, the latter only describes the deformation at the linearized level. A generalization to all orders in the deformation was proposed in \cite{Benitez} in the case when the deformation satisfies the Yang-Baxter equation. 

In this paper we derive the supergravity solution of Maldacena and Lunin from the worldsheet pure spinor sigma model of the beta deformation. We then proceed to examine the beta deformed theory in the fishnet limit, where we identify a degeneration in the background fields. We explain how the BRST structure of the theory imposes restriction on the structure of degeneration.
The space of all possible degenerate limits of a sigma-model is very rich. But, BRST invariance implies that the degeneration is of a very special, ``regular" kind.

\subsection{Supergravity description}

The $\beta-$deformed $\mathcal{N}=4$ SYM theory is related via gauge-gravity duality to a supergravity solution which is a deformation of the standard $AdS_5\times S^5$ one. This deformation only acts on the $S^5$ geometry and leaves the $AdS_5$ part intact. This solution was obtained by Maldacena, Lunin and Frolov by compactifying M theory on a torus of modulus $\tau$ \cite{Lunin_2005},\cite{Frolov_2005}. The parameter $\tau$ is acted upon by the $SL(2; \mathbb{R})$ symmetry of the supergravity equations of motion. Then, the beta deformation of $\mathcal{N}=4$ SYM with parameter $\gamma$ corresponds to the transformation
\begin{equation}\label{tau}
\tau \rightarrow \frac{\tau}{1+\gamma\tau},
\end{equation}
which acts as a solution generating transformation. This action is obtained from a T-duality on one of the torus circles, then a shift of coordinates, and then another T-duality. Such procedure is called $TsT$. More precisely, the solution found by Maldacena-Lunin-Frolov is given by the following metric and $B$-field:
\begin{equation}\label{MaldLun}
ds^2 = R^2\left[ ds^2_{AdS_5} +  \sum_{i=1}^3(d\mu_i^2 + G\mu_i^2d\phi_i^2) + G\mu_1^2\mu_2^2\mu_3^2\left(\sum_{i=1}^3\hat\gamma_id\phi_i\right)^2 \right],
\end{equation}
\begin{equation}
B = R^2G(\hat\gamma_3\mu_1^2\mu_2^2d\phi_1\wedge d\phi_2+\hat\gamma_1\mu_2^2\mu_3^2d\phi_2\wedge d\phi_3+\hat\gamma_2\mu_3^2\mu_1^2d\phi_3\wedge d\phi_1)\nonumber,
\end{equation}
with $\hat\gamma_i = R^2\gamma_i$ and $G^{-1} = 1 + \hat\gamma_3^2\mu_1^2\mu_2^2+\hat\gamma_1^2\mu_2^2\mu_3^2+\hat\gamma_2^2\mu_3^2\mu_1^2$. As we can see, this corresponds to a deformation in the sphere metric, which is described by $ds^2=R^2\sum_{i=1}^3(d\mu_i^2 + \mu_i^2d\phi_i^2)$, with $\sum_{i=1}^3\mu_i^2=1$ and $R$ the radius of the sphere. In the limit of $\gamma_i=0$ we recover the sphere metric and then the $AdS_5\times S^5$ undeformed solution.

\subsection{Deriving supergravity from worldsheet pure spinor description}

One of the results of this paper is the explicit demonstration that the sigma model proposed in \cite{Benitez} reproduces the supergravity solution of Maldacena-Lunin-Frolov. The first step to obtain this model is the insertion of a vertex operator in the $AdS_5\times S^5$ superstring action \cite{andrei}. However, this procedure only provides a description of the linearized beta deformation. In order to include the non-linear behaviour, one must impose BRST invariance of the action at all orders in the deformation parameter. The reason why this computation is possible is that the expansion of the BRST operator in powers of deformation parameter stops at the first order.

We can introduce embedding coordinates by the embedding $AdS_5\times S^5\subset\mathbb{R}^6\times\mathbb{R}^6$. 
Doing this, we can parametrize the space-time in an easier way and then write the deformed sigma model action. Using the 
coordinate system derived from this embedding, the resulting sigma model is
\begin{equation}
S_b = \int d^2z\ (g_{mn}+B_{mn})\partial x^m\bar\partial x^n,
\end{equation}
where the background fields $g_{mn}$ and $B_{mn}$ are precisely the metric and the $B$-field in the supergravity solution (\ref{MaldLun}), and $x^m$ denotes the bosonic space-time coordinates of $AdS_5\times S^5$.

We therefore are able to rederive the supergravity solution obtained from TsT in \cite{Lunin_2005,Frolov_2005} using string theory techniques.

\subsection{Large $\gamma$}
Another aim of the paper is to work towards a string dual description of the fishnet field theories \cite{Gurdogan:2015csr}. In order to do this, we must work with the fully deformed theory since it allows us to study the sigma model in the strong deformation regime. These theories are obtained from the beta deformed $\mathcal{N}=4$ SYM in a particular Double Scaling limit. This limit is given by taking $\gamma\rightarrow\infty$, $\lambda\rightarrow 0$ with $\lambda e^{-i\frac{\gamma}{2}}$ fixed, where $\gamma$ is the deformation parameter and $\lambda$ the t'Hooft coupling constant. To study this double scaling limit in the gravity side, we must take into account the relation between t'Hooft constant and the AdS radios given by $\lambda = R^4/\alpha'^2$, and then impose the fishnet limits
\begin{align}
\gamma_i\rightarrow i\infty\ ,\ \ \ \frac{R}{\sqrt{\alpha'}}\rightarrow 0,\\
R e^{-i\frac{\gamma_i}{2}} = \xi_i = \text{const}.\nonumber
\end{align}
In principle, we must impose the condition $\gamma << 1$ in the supergravity solution (\ref{MaldLun}). The reason is that the transformation (\ref{tau}) in the torus $\tau \rightarrow \frac{\tau}{1+\gamma \tau}$ cannot make the torus smaller than the string scale. Also, we must have $R>>1$. Then, it is clear that in order to study the fishnet limit we must use the fully deformed theory, since it is defined at all orders in $\gamma$. This is the precise advantage of our proposal. We must emphasize that, in this paper, we study the large $\gamma$ limit but with constant radius $R$. This would be the first step to understanding the string dual for the fishnet theories. The small radius limit has not been investigated in this context yet.

The BRST operator of the theory in the strong deformation regime is given by
\begin{equation}
Q = B^{ab}\Lambda_bt_a.
\end{equation}
The indices $a$ are bosonic indices of the algebra of rotations $\mathfrak{su}(4)\subset\mathfrak{psu}(2,2|4)$, $\Lambda_a$ are functions of the sigma model fields, $t_a$ are generators of $\mathfrak{su}(4)$ and $B^{ab}$ an antisymmetric constant matrix. The BRST invariance of the model will imply a degeneracy of the background fields in the $t_a$ directions,
\begin{equation}
G^f_{\mu\nu}t_a^\nu =0,\ \ \ B^f_{\mu\nu}t_a^\nu =0\;,
\end{equation}
where $G^f_{\mu\nu}$ and $B^f_{\mu\nu}$ are the metric and the B-field of the theory in the fishnet limit.

\subsection{Plan of the paper}

In section 2 we review the Pure Spinor sigma model in $AdS_5\times S^5$. In section 3 we construct the beta deformed theory in all orders in the deformation parameter. There, we write the action for the full beta deformation, as well as the BRST operator. We then compute explicitly the bosonic part of our action and find a complete agreement with the Maldacena-Lunin-Frolov supergravity solution. In section 4 we study the fishnet limit applied to our sigma model and show that the background fields degenerate in this regime.

\section{Review of \texorpdfstring{$AdS_5\times S^5$ } SSuperstring: pure spinor formulation}

In the superstring in the $AdS_5\times S^5$ background, the fields are maps from the world-sheet $\Sigma$ to the following coset:
\begin{equation}
g:\Sigma\longrightarrow \frac{PSU(2,2|4)}{SO(4,1)\times SO(5)}
\end{equation}

Considering $(\tau^+,\tau^-)$ the parametrization of the worldsheet, we will write the action in terms of the invariant current $J=J_+d\tau^++J_-d\tau^-$, where $J_{\pm}=-\partial_{\pm}gg^{-1}$ and $\partial_{\pm}=\frac{\partial}{\partial\tau^\pm}$. One decomposes the lie algebra in the following way: $\mathfrak{psu}(2,2|4)=\mathfrak{g}_0\oplus\mathfrak{g}_1\oplus\mathfrak{g}_2\oplus\mathfrak{g}_3$, with $\mathfrak{g}_0=\mathfrak{so}(4,1)\oplus\mathfrak{so}(5)$. The massive part of the action can be written in terms of the currents $J_1,J_2,J_3$. In the pure spinor formalism \cite{Berkovits_2000}, the whole action is 
\begin{equation}\label{S0}
S_0 = \frac{R^2}{2\alpha'}\int \Str \Big( \frac{1}{2}J_{2+}J_{2-}+\frac{3}{4}J_{1+}J_{3-} + \frac{1}{4}J_{3+}J_{1-}\nonumber
\end{equation}
\begin{equation}
\hspace{4cm} + w_{1+}D_{-}\lambda_3+ w_{3-}D_{+}\lambda_1 - N_{0+}N_{0-} \Big).
\end{equation}
It is remarkable here that we introduced two ghost fields $\lambda_1,\lambda_3$, which are respectively right and left \textit{pure spinors}. These are spinor fields with values in the lie algebra, satisfying the pure spinor constraints characterized by 
\begin{equation}\label{psconstraint}
\{\lambda_1,\lambda_1\}=\{\lambda_3,\lambda_3\}=0.
\end{equation}
The fields $w_{1+}$ and $w_{3-}$ are respectively the conjugate momenta associated to $\lambda_3$ and $\lambda_1$. $N_0$ are the Lorentz currents, defined as $N_{0+}=-\{w_{1+},\lambda_3\}$, $N_{0-}=-\{w_{3-},\lambda_1\}$ and $D_{\pm}=\partial_\pm + [J_{0\pm},-]$ is the covariant derivative. In terms of the algebraic structure underlying the ghost fields, $\lambda_1$ and $w_{1+}$ are elements of $\mathfrak{g}_1$, while $\lambda_3$ and $w_{3-}$ are elements of $\mathfrak{g}_3$.

The introduction of the ghost sector is crucial for achieving covariant quantization of the superstring, as discussed in \cite{Berkovits_2000}. This accomplishment is realized through the definition of a nilpotent BRST-like operator, parameterized by the pure spinor ghosts, as:
\begin{align}
&Q_0\lambda_1=Q_0\lambda_3=0\\
& \epsilon Q_0g = (\epsilon\lambda_3+\epsilon\lambda_1)g\\
& \epsilon Q_0(w_{1+}) = -\epsilon J_{1+},\ \ \ \epsilon Q_0(w_{3-}) = -\epsilon J_{3-}.
\end{align}

The constraints (\ref{psconstraint}) that the ghost fields satisfy imply a non-trivial cohomology for this BRST-like operator. Consequently, it is natural to define the physical states of the system as elements within this cohomology, called vertex operators. A particularly notable example is the construction of the zero-mode dilaton vertex in \cite{Mikhailov_2012,Berkovits_2009}, which is defined by
\begin{equation}\label{dil}
V=\Str(\lambda_1\lambda_3).
\end{equation}

This vertex can be seem as an state resulting from the rescaling of the overall string coupling constant. Thus, in essence, the string theory is deformed by an expression given by the integral of the Lagrangian itself. The action is invariant under BRST transformation and the Lagrangian cannot be BRST-exact, therefore it is in the cohomology. Then, in order to find the corresponding unintegrated vertex operator, writen in (\ref{dil}), one uses that the BRST variation of the Lagrangian is $Q_0 \mathcal{L}_0 = d\Str(\lambda_3J_1-\lambda_1J_3)$, which results in the following descent equations:
\begin{align}
&Q_0 \mathcal{L}_0 = d\Big(\Str(\lambda_3J_1-\lambda_1J_3)\Big),\\
&Q_0 \Big(\Str(\lambda_3J_1-\lambda_1J_3)\Big) = d\Str(\lambda_1\lambda_3) .
\end{align}

In the next section, after a modification of this vertex operator, it will play a crucial rule in the study of the beta deformation.

We should now define an important quantity for the superstring, the density of the global conserved charges, given by $j = j_+dz^++j_-dz^-$, where
\begin{align}\label{j}
&j_+ = -g^{-1}( J_{3+} + 2J_{2+} + 3J_{1+} + 4N_{0+}  )g\\
&j_- = g^{-1}( J_{1-} + 2J_{2-} + 3J_{3-} + 4N_{0-}  )g,\nonumber
\end{align}
such that $dj=0$. The density current transforms under BRST action as
\begin{equation}\label{descent}
\epsilon Q(j) = d\Lambda(\epsilon),
\end{equation}
where $\Lambda(\epsilon) = g^{-1}\epsilon(\lambda_3-\lambda_1)g$. For any element of the Lie algebra $x\in\mathfrak{g}$, we can write the components of $x$ by
\begin{equation}
x_a = \Str(xt_a),
\end{equation}
where $\{t_a\}$ are the generators of the lie algebra $\mathfrak{g}$.

\section{$\beta$-deformation of the superstring sigma model}

In this section, we will first review the beta deformation of the pure spinor superstring, as introduced in \cite{andrei} 
and \cite{Benitez}. We will review how to obtain the worldsheet action and the BRST operator in the presence of this deformation. After the review, we will establish a new connection between this worldsheet action and the 
Maldacena-Lunin construction in \cite{Lunin_2005}.

To start with, one considers a modification of the dilaton vertex (\ref{dil}). This modification was proposed in \cite{andrei} in order to describe the beta deformation in string theory and is given by the following expression:
\begin{equation}\label{jj0}
V^{(0)}_{ab}=(g^{-1}(\lambda_3-\lambda_1)g)_a(g^{-1}(\lambda_3-\lambda_1)g)_b = \Lambda_a\Lambda_b.
\end{equation}
This vertex will result in an integrated vertex operator given by the product of two conserved currents,
\begin{equation}\label{jj}
V^{(2)}_{ab} =  j_{[a}\wedge j_{b]}.
\end{equation}
As we will see in the end of this section, this vertex operator gives the supergravity solutions corresponding 
to the beta deformation, described in \cite{Lunin_2005}.

The expression for the integrated vertex in (\ref{jj}) is obtained from (\ref{jj0}) via the following descent equations, computed by using the relation in (\ref{descent}):
\begin{align}
& d\Big(\Lambda_a\Lambda_b\Big) = Q \Big(j_{[a}\Lambda_{b]} \Big),\\
& d\Big(j_{[a}\Lambda_{b]}\Big) = Q\Big(j_{[a}\wedge j_{b]}\Big).
\end{align}

The $AdS_5\times S^5$ superstring is deformed by adding the integrated vertex (\ref{jj}) to its Lagrangian. The deformed action is then
\begin{equation}
S_{def} = S_0 +  \frac{\gamma}{4}\int B^{ab}j_{[a}\wedge j_{b]},
\end{equation}
where $B^{ab}$ is an antisymmetric matrix with two adjoint indices. This is a linearized deformation of the action in the parameter $\gamma$. The next step now is to think about the second order deformation in $\gamma$. To do that, a deformation is added to the BRST operator,
\begin{equation}\label{qdef}
Q_{def} = Q_0 + \gamma Q_1,
\end{equation}
in such a way that the action remains BRST invariant, which is guaranteed  by the equation
\begin{equation}\label{consistence}
Q_1 S_0 + Q_0\int V_1^{(2)}=0.
\end{equation}
By acting with $Q_0$ on $V_1^{(2)}$ one gets $Q_0\int V_1^{(2)}=-\frac{1}{2}\int B^{ab}j_a\wedge d\Lambda_b$. Then, equation (\ref{consistence}) implies that we must find a $Q_1$ such that
\begin{equation}\label{canc}
Q_1S_0 =\frac{1}{2}\int B^{ab}j_a\wedge d\Lambda_b.
\end{equation}
In order to find $Q_1$, one notice that the deformation of the BRST operator acting on $g$ is a $\mathfrak{psu}(2,2|4)$ transformation with a space-time dependent parameter $\alpha(z^+,z^-)$, $Q_1g= g\alpha$. By using this in equation (\ref{canc}) one finds the following expression for the deformation of the BRST operator at first order:
\begin{equation}\label{brst}
Q_1g = gB^{ab}\Lambda_at_b.
\end{equation}
The actions on the $w$ ghost were also computed in \cite{Benitez} in a similar manner. We then write the first 
order correction for the BRST operator as
\begin{align}
&Q_1g = gB^{ab}\Lambda_at_b\\
& Q_1w_{1+}=\big(B^{ab}j_{a+}(gt_bg^{-1})\big)_1,\ \ \ Q_1w_{3-}=\big(B^{ab}j_{a-}(gt_bg^{-1})\big)_3.
\end{align}

\subsection{Simplification of the BRST operator}

The deformation of the BRST operator we are studying is hugely simplified when the matrix $B$ satisfy the following Yang-Baxter equation:
\begin{equation}\label{YB}
[\![ B,B ]\!]^{abc} \equiv B^{p[a}f^{\ \ b}_{pq}B^{c]q} = 0.
\end{equation}
Namely, under this condition the BRST operator acting on $g$ is deformed only up to the first order in $\gamma$. From this condition it also follows that $Q_1^2g=0$ automatically. Indeed,

\begin{align}
Q_1^2 g & = \Lambda_aB^{ab}\mathcal{L}_{t_b}(\Lambda_cB^{cd}\mathcal{L}_{t_d} g)\nonumber\\
& = \Lambda_aB^{ab}f_{bc}^{\ \ e}\Lambda_eB^{cd}\mathcal{L}_{t_d}g + \frac{1}{2}\Lambda_aB^{ab}\Lambda_cB^{cd}f_{bd}^{\ \ e}\mathcal{L}_{t_e}g\\
& = \frac{1}{2}B^{p[a}f^{\ \ b}_{pq}B^{c]q}\Lambda_a\Lambda_b\mathcal{L}_{t_c}g=0.\nonumber
\end{align}
Now, since $Q_0$ and $Q_1$ commute, the operator given by $Q_0+\gamma Q_1$ is nillpotent by itself. Then, in conclusion, the correction for the BRST operator acting on the group element stops at first order, and the full operator acting on $g$ is then just $Q_{def}=Q_0+\gamma Q_1$. 

\subsection{Deformation to all orders}

In order to find the fully deformed Lagrangian, one imposes the BRST invariance of the action to all orders in $\gamma$. From 
the BRST invariance condition follow an infinite set of equations, one for each order in $\gamma$,
\begin{equation}\label{consistency}
Q_nS_0+\sum_{i=1}^n Q_i\int V_{n-i}^{(2)}=0,\ \ \forall n.
\end{equation}
The equation corresponding to the first order in $\gamma$ is $Q_1S_0+Q_0\int V_1=0$ and it was explained in the beginning of this section. The equation for the second order in $\gamma$ is $Q_2S_0+Q_1\int V_1+Q_0\int V_2=0$, from which one obtains expressions for $V_2$ and $Q_2$. This was systematically computed in \cite{Benitez}. Finally, the full deformation for the action is
\begin{equation}
\delta S = \int\ \langle j,\kappa j\rangle ,\ \ \ \text{where}\ \ 
\kappa = \gamma B\cdot\sum_{n=0}^\infty \Big(\gamma M\cdot B\Big)^n = \gamma B\cdot\left(1-\gamma MB\right)^{-1}\;,
\end{equation}
where $j$ is the conserved current (\ref{j}) and $M= \Str\Big[(gt_ag^{-1})d_{PS}(gt_bg^{-1})\Big]$, with $d_{PS} = P_1+2P_2+3P_3$\footnote{This can be written explicitly as $M_{ab} = \Str\Big[(gt_ag^{-1})_1(gt_bg^{-1})_3+2(gt_ag^{-1})_2(gt_bg^{-1})_2+3(gt_ag^{-1})_3(gt_bg^{-1})_1\Big]$.}.
And the full correction to the BRST operator is
\begin{equation}
\delta Q = \gamma B^{ab}\Lambda_at_b + \left(\kappa^{ab} j_{a+}(gt_bg^{-1})\right)_1\frac{\delta}{\delta w_{1+}} + \left(\kappa^{ab} j_{a-}(gt_bg^{-1})\right)_3\frac{\delta}{\delta w_{3-}}.
\end{equation}
We then write the full deformation as
\begin{equation}\label{kappa}
\delta S = \int\ \kappa^{ab}_\gamma j_{a+} j_{b-},\ \ \text{with}\ \  \kappa_\gamma^{ab}=\gamma B^{ac} \left[(1-\gamma MB)^{-1}\right]_c^{\ b}\;,
\end{equation}
and then we obtain our full sigma model,

\begin{equation}\label{defaction}
S_{def} = \frac{R^2}{2\alpha'}\int \Str \Big( \frac{1}{2}J_{2+}J_{2-}+\frac{3}{4}J_{1+}J_{3-} + \frac{1}{4}J_{3+}J_{1-}+ \kappa^{ab}_\gamma j_{a+} j_{b-} \nonumber
\end{equation}
\begin{equation}\label{def}
\hspace{4cm} + w_{1+}D_{-}\lambda_3+ w_{3-}D_{+}\lambda_1 - N_{0+}N_{0-}  \Big)\;,
\end{equation}
with the BRST operator given by
\begin{align}\label{defQ}
Q =\ &(\lambda_1+\lambda_3)g\frac{\delta}{\delta g} -J_{1+}\frac{\delta}{\delta w_{1+}}-J_{3-}\frac{\delta}{\delta w_{3-}}\nonumber\\
&+\gamma gB^{ab}\Lambda_at_b\frac{\delta}{\delta g} + \left(\kappa^{ab} j_{a+}(gt_bg^{-1})\right)_1\frac{\delta}{\delta w_{1+}} + \left(\kappa^{ab} j_{a-}(gt_bg^{-1})\right)_3\frac{\delta}{\delta w_{3-}}.
\end{align}

In the next sections, we will describe the connection between the worldsheet action in (\ref{defaction}), obtained in \cite{Benitez}, with the Maldacena-Lunin construction in \cite{Lunin_2005}.

\subsection{Maldacena-Lunin background}

The Maldacena-Lunin background \cite{Lunin_2005} is a supergravity dual of the beta deformed field theory. In order to obtain it from our sigma model we will restrict to the bosonic part and only deform the 5-sphere, letting $AdS_5$ undeformed. The sphere can be parametrized by the embedding coordinates
\begin{align}\label{coord}
&u_1+iu_2 =\mu_1 e^{i\phi_1},\nonumber \\
&u_3+iu_4 =\mu_2 e^{i\phi_2},\nonumber \\
&u_5+iu_6 =\mu_3 e^{i\phi_3},,\ \ \ \ \mu_1^2+\mu_2^2+\mu_3^2=1.
\end{align}

Different choices of the matrix $B$ in 
(\ref{kappa}) will give us different backgrounds. Let us chose a $B$ such that its only non-vanishing components  are
\begin{equation}\label{B}
B^{[12][34]} = B^{[34][56]} = B^{[56][12]} = 1\;,
\end{equation}
where $ij$ are indices for the $\mathfrak{su}(4)$ subgroup of $\mathfrak{psu}(2,2|4)$ (see Appendix \ref{emb}). This $B$ satisfy the Yang-Baxter equation (\ref{YB})
\begin{equation}
[\![B,B]\!]^{[ij][mn][kl]} = B^{[ij][mp]}B^{[pn][kl]} - B^{[ij][pn]}B^{[mp][kl]} = 0.
\end{equation}

In this section, we will compute the bosonic part of the matrix $\kappa$ in (\ref{kappa}) and then write the action in therms of the embedding coordinates (\ref{coord}). Consider the vector space $\mathfrak{su}(4)$ given by the basis $t_{ij}$ with $i,j$ anti-symmetric going from $1$ to $6$. Ordering the basis as $\{ t_{12},t_{34},t_{56},\cdots \}$, we have
\begin{equation}
B = \left[
\begin{array}{ccc;{2pt/2pt}cccc}
0 & 1 & -1 & \ & \ & \ & \ \\
-1 & 0 & 1 & \ & \ 0 & \ & \ \\
1 & -1 & 0 & \ &  & \ & \ \\ \hdashline[2pt/2pt]
\ & \ & \ & \ &  & \ & \ \\
\ & 0 & \ & \ & \ 0 & \ & \ \\
\ & \ & \ & \ & \ & \ & \ \\
\end{array}
\right].
\end{equation}

From now on, we will restrict our vector space to a $3$-dimensional space generated by these three elements $\{t_{12},t_{34},t_{56}\}$. Then, we will write $B$ as a $3\times 3$ matrix
\begin{equation}\label{Bm}
B = \left(
\begin{array}{ccc}
 0 & 1 & -1 \\
 -1 & 0 & 1 \\
 1 & -1 & 0 \\
\end{array}
\right)\;,
\end{equation}
as well as the matrix $M$
\begin{equation}\label{M}
M=\left(
\begin{array}{ccc}
M_{12,12} & M_{12,34} & M_{12,56} \\
M_{34,12} & M_{34,34} & M_{34,56} \\
M_{56,12} & M_{56,34} & M_{56,56} \\
\end{array}
\right).
\end{equation}

\subsubsection{The structure of $M_{ab}$}
The matrix $M$ is rather complicated,
\begin{equation}
M_{ab} = \Str\Big[ (gt_ag^{-1})_1(gt_bg^{-1})_3 + 2(gt_ag^{-1})_2(gt_bg^{-1})_2 + 3(gt_ag^{-1})_3(gt_bg^{-1})_1  \Big]
\end{equation}
It may be decomposed into a symmetric and antisymmetric part,
\begin{align}
&M_{(ab)}= 2\Str\Big[ (gt_ag^{-1})_1(gt_bg^{-1})_3 + (gt_ag^{-1})_2(gt_bg^{-1})_2 + (gt_ag^{-1})_3(gt_bg^{-1})_1  \Big]\\
&M_{[ab]} = \Str\Big[ (gt_ag^{-1})_3(gt_bg^{-1})_1 - (gt_ag^{-1})_1(gt_bg^{-1})_3 \Big].
\end{align}
Using the embedding bosonic coordinates $(u_1,\cdots,u_6)$ in (\ref{coord}), we may compute explicitly the matrix $M$ restricted to the bosonic coordinates. In this case, as shown in equation (\ref{mij}) of the Appendix \ref{emb}, the only non-vanishing components for $M$ are $M_{[12][12]}, M_{[34][34]}, M_{[56][56]}$,
\begin{equation}\label{m2}
M^{(2)}=\left(
\begin{array}{ccc}
 u_{1}^2+u_{2}^2 & 0 & 0 \\
 0 & u_{3}^2+u_{4}^2 & 0 \\
 0 & 0 & u_{5}^2+u_{6}^2 \\
\end{array}
\right)=\left(
\begin{array}{ccc}
 \mu_{1}^2 & 0 & 0 \\
 0 & \mu_{2}^2 & 0 \\
 0 & 0 & \mu_{3}^2 \\
\end{array}
\right).
\end{equation}
The bosonic action will be given by 
\begin{equation}\label{Sb}
S_b^\gamma = \frac{1}{2}\Str[J_2\bar J_2] + \langle j,\kappa_\gamma^{(2)}j\rangle,
\end{equation}
where
\begin{equation}\label{kappa2}
\kappa_\gamma^{(2)}=\left(B \frac{\gamma}{1-\gamma M^{(2)}B}\right).
\end{equation}
If we substitute the expression for $M^{(2)}$ given by equation (\ref{m2}) into the matrix $\kappa_\gamma^{(2)}$ in (\ref{kappa2}) and use the matrix $B$ in (\ref{Bm}), we find 
\begin{equation}
\kappa^{(2)}_\gamma = \frac{\gamma^2}{G}\times\left(
\begin{array}{ccc}
 -(\mu_2 ^2+\mu_3 ^2) &    \mu_3 ^2& \mu_2 ^2 \\
\mu_3^2 & -(\mu_1 ^2+\mu_3 ^2) &  \mu_1 ^2 \\
\mu_2 ^2 & \mu_1 ^2 & - \left(\mu_1^2+\mu_2 ^2\right) \\
\end{array}
\right)+\frac{\gamma}{G}\times\left(
\begin{array}{ccc}
0 & 1 & -1 \\
-1 & 0 & 1 \\
1 &  -1 & 0 \\
\end{array}
\right)\;,
\end{equation}
where $G =1+ \gamma ^2 \left(\mu_1 ^2 \mu_2 ^2+\mu_1 ^2 \mu_3 ^2+\mu_2 ^2 \mu_3 ^2\right)$. The currents $j$, also computed in Appendix \ref{emb}, equation (\ref{j2}), can be written as $j_{12} = \mu_1^2d\phi_1$, $j_{34} = \mu_2^2d\phi_2$ and $j_{56} = \mu_3^2d\phi_3$. We use this expression for the currents and the matrix $\kappa^{(2)}_\gamma$ above to compute the bosonic action (\ref{Sb}). The result is 

\begin{equation}\label{beta}
S_b^\gamma = S_{AdS_5}+R^2\int d^2z\ \Bigg[ \sum_{i=1}^3(d\mu_i\wedge\star d\mu_i + \frac{\mu^2_i}{G}d\phi_i\wedge\star d\phi_i) + \frac{\gamma^2}{G}\mu^2_1\mu^2_2\mu^2_3\left(\sum_{i=1}^3d\phi_i\right)\wedge\star\left(\sum_{i=1}^3d\phi_i\right)\nonumber
\end{equation}
\begin{equation}
\hspace{5cm} + \frac{\gamma}{G}\left( \mu_1^2\mu_2^2\ d\phi_1\wedge d\phi_2+\mu_2^2\mu_3^2\ d\phi_2\wedge d\phi_3+\mu_3^2\mu_1^2\ d\phi_3\wedge d\phi_1 \right)\Bigg].
\end{equation}
This gives us the metric and the $B$-field of the sigma model,
\begin{equation}
g_\gamma = \sum_{i=1}^3\left(d\mu_i^2 + G^{-1}\mu_i^2d\phi_i^2\right) + G^{-1}\gamma^2\mu_1^2\mu_2^2\mu_3^2\left(\sum_{i=1}^3d\phi_i\right)^2
\end{equation}
\begin{equation}
B = G^{-1}\gamma\left( \mu_1^2\mu_2^2\ d\phi_1\wedge d\phi_2+\mu_2^2\mu_3^2\ d\phi_2\wedge d\phi_3+\mu_3^2\mu_1^2\ d\phi_3\wedge d\phi_1 \right)\;,
\end{equation}
which is the Maldacena-Lunin solution \cite{Lunin_2005}.

\subsection{Generalization to $\gamma_i$ deformation}\label{frolov}

In \cite{Frolov_2005}, Frolov studied the supergravity dual to a generalization of the beta deformation by considering 3 parameters $\gamma_i$. This can be captured in our sigma model by labeling the deformation parameter as
\begin{equation}
\gamma_i = \alpha_i\gamma,
\end{equation}
with constants $\alpha_i$, for $i=1,2,3$. Then, we define our theory based on the following matrix $B$:
\begin{equation}\label{gammaB}
B = \left(
\begin{array}{ccc}
 0 & \alpha _3 & -\alpha _2 \\
 -\alpha _3 & 0 & \alpha _1 \\
 \alpha _2 & -\alpha _1 & 0 \\
\end{array}
\right)\;,
\end{equation}
obtained by setting $B^{[12][34]}=\alpha_3$, $B^{[34][56]}=\alpha_1$ and $B^{[56][12]}=\alpha_2$. The matrix $\kappa$ 
becomes

\begin{equation}\label{kappagammai}
\kappa_{\gamma_i}^{(2)} =\gamma B\frac{1}{1-\gamma M^{(2)}\cdot B}.
\end{equation}

Now, we use the expression for $M^{(2)}$ we computed in (\ref{m2}) and the matrix $B$ defined above in (\ref{gammaB}) to compute $\kappa_{\gamma_i}^{(2)}$ in (\ref{kappagammai}). As a result we find 

\begin{equation}
\kappa_{\gamma_i}^{(2)} = \frac{1}{G}\left(
\begin{array}{ccc}
 -\gamma _3^2 \mu _2^2-\gamma _2^2 \mu _3^2 & \gamma _1 \gamma _2 \mu _3^2 & \gamma _1 \gamma _3 \mu _2^2 \\
 \gamma _1 \gamma _2 \mu _3^2 & -\gamma _3^2 \mu _1^2-\gamma _1^2 \mu _3^2 & \gamma _2 \gamma _3 \mu _1^2 \\
 \gamma _1 \gamma _3 \mu _2^2 & \gamma _2 \gamma _3 \mu _1^2 & -\gamma _2^2 \mu _1^2-\gamma _1^2 \mu _2^2 \\
\end{array}
\right) + \frac{1}{G}\left(
\begin{array}{ccc}
 0 & \gamma _3 & -\gamma _2 \\
 -\gamma _3 & 0 & \gamma _1 \\
 \gamma _2 & -\gamma _1 & 0 \\
\end{array}
\right)\;,
\end{equation}
where $G = 1 +\gamma _3^2 \mu _1^2 \mu _2^2+\gamma _1^2 \mu _3^2 \mu _2^2+\gamma _2^2 \mu _1^2 \mu _3^2$. The corresponding sigma model can be computed using this result and the expressions for the currents $j_{12},j_{34},j_{56}$ applied to equation (\ref{Sb}). The result is the bosonic action

\begin{equation}\label{Sfrolov}
S_b^{\gamma_i} = S_{AdS_5}+R^2\int d^2z\ \Bigg[ \sum_{i=1}^3 (d\mu_i\wedge\star d\mu_i + \frac{\mu^2_i}{G}d\phi_i\wedge\star d\phi_i) + \frac{1}{G}\mu^2_1\mu^2_2\mu^2_3\left(\sum_{i=1}^3\gamma_id\phi_i\right)\wedge\star\left(\sum_{i=1}^3\gamma_id\phi_i\right)\nonumber
\end{equation}
\begin{equation}
\hspace{3cm} + \frac{1}{G}\left( \gamma_3\ \mu_1^2\mu_2^2\ d\phi_1\wedge d\phi_2+\gamma_1\ \mu_2^2\mu_3^2\ d\phi_2\wedge d\phi_3+\gamma_2\ \mu_3^2\mu_1^2\ d\phi_3\wedge d\phi_1 \right)\Bigg].
\end{equation}
This gives us the metric and the $B$-field of the sigma model already described by Frolov \cite{Frolov_2005},
\begin{equation}
g_\gamma = \sum_{i=1}^3\left(d\mu_i^2 + G^{-1}\mu_i^2d\phi_i^2\right) + G^{-1}\mu_1^2\mu_2^2\mu_3^2\left(\sum_{i=1}^3\gamma_id\phi_i\right)^2
\end{equation}
\begin{equation}
B = G^{-1}\left( \gamma_3\mu_1^2\mu_2^2\ d\phi_1\wedge d\phi_2+\gamma_1\mu_2^2\mu_3^2\ d\phi_2\wedge d\phi_3+\gamma_2\mu_3^2\mu_1^2\ d\phi_3\wedge d\phi_1 \right).
\end{equation}

 \subsection{Symmetries}\label{symmetries}

The superconformal transformation of the current $j$ and the matrix $M$ is:
\begin{equation}
\delta_a j_b = f_{ab}^{\ \ c}j_c
\end{equation}
\begin{equation}
\delta_e M_{ab} = f^{\ \ c}_{ea}M_{cb}+f^{\ \ c}_{eb}M_{ac}.
\end{equation}

The currents $j_{12},j_{34},j_{56}$ are all invariant under the plane rotations $t_{12},t_{34},t_{56}$. This happens because all these generators commutes among themselves, since the intersection of the plane defined by then is just a point. Therefore the structure constants $f_{ab}^{\ \ I}$ are all zero for $a,b\in \{12, 34, 56\}$. For example, $f^{\ \ \ \ ij}_{12,34}=\delta_{[i|[4}\delta_{3][2}\delta_{1]|j]}-\delta_{[i|[2}\delta_{1][4}\delta_{3]|j]}=0$. In general we have
\begin{align}
&t_{a}j_{b} = f^{\ \ I}_{ab} j_{I} = 0,\ \ a,b\in \{12, 34, 56\}.
\end{align}
The same thing happens for the variation of the components $M_{ab}$ with indices $a,b\in \{12, 34, 56\}$, in such a way that they are also invariant under these three rotations,
\begin{align}
&t_{a} M_{bc} = f_{ab}^{\ \ \ I}M_{I,c}+f_{ac}^{\ \ \ \ I}M_{b,I} = 0.
\end{align}
The action for the beta deformation was constructed using only the components for the currents $j_{12},j_{34},j_{56}$ and $M_{ab}$ with indices $a,b\in \{12, 34, 56\}$. Therefore, the action is invariant under $U(1)^3$. This is a residual symmetry from the isometry group of the sphere $SU(4)$. This also happens in the field side \cite{Gurdogan:2015csr}. This symmetry will important later in the next section.

\section{The large $\gamma$ limit}\label{fishnet}

In this section, we will study the worldsheet action under strong deformations. This is motivated by the work of 
\cite{Gurdogan:2015csr}, where they studied a field theory in the limit of strong deformations. Specifically, the 
$\mathcal{N}=4$ SYM theory is $\gamma_i$-deformed and a double scaling limit is taken, giving rise to a field theory 
named \textit{fishnet theory}. This limit is characterized by the imaginary part of the deformation parameter $\gamma_j$ 
being infinitely large ($\gamma_j\rightarrow i\infty$), the coupling constant $g^2=Ng_{YM}^2$ being zero ($g^2\rightarrow 0$), 
and $\xi_j^2=g^2e^{-i\gamma_j}$ being constant, where $j=1,2,3$. This section makes a similar analysis for the 
worldsheet, in order to establish a possible candidate for the string dual of this fishnet field theory.

For simplicity, one can start studying the strong deformed sigma model by setting two deformation parameters to zero, 
say $\gamma_2$ and $\gamma_3$, and keep the remaining $\gamma_1$ nonzero. In this case, the bosonic part of the 
sigma model action is 
\begin{equation}\label{g1}
S_b^{\gamma_1} = S_{AdS_5}+R^2\int d^2z\ \Bigg[ \sum_{i=1}^3 d\mu_i\wedge\star d\mu_i + \frac{\mu^2_i}{1 +\gamma _1^2 \mu _3^2 \mu _2^2}d\phi_i\wedge\star d\phi_i + \frac{\gamma^2_1\mu^2_1\mu^2_2\mu^2_3}{1 +\gamma _1^2 \mu _3^2 \mu _2^2}d\phi_1\wedge\star d\phi_1\nonumber
\end{equation}
\begin{equation}
\hspace{3cm} + \frac{\gamma_1\mu_2^2\mu_3^2}{1 +\gamma _1^2 \mu _3^2 \mu _2^2} d\phi_2\wedge d\phi_3\Bigg].
\end{equation}
Then, in the limit of strong deformation, $\gamma_1\rightarrow i\infty$, we have
\begin{equation}\label{f1}
S_b^{\gamma_1\rightarrow i\infty} = S_{AdS_5}+R^2\int d^2z\ \Bigg[ \sum_{i=1}^3 d\mu_i\wedge\star d\mu_i + \mu_1^2d\phi_1\wedge\star d\phi_1\Bigg],
\end{equation}
which makes manifest a degeneracy in the action. This may reflect the decoupling of some fields already present in the 
fishnet field theory \cite{Gurdogan:2015csr}. In the general case where three deformation parameters are considered, 
the bosonic action assumes the form

\begin{equation}\label{f123}
S_b^{\gamma_1,\gamma_2,\gamma_3\rightarrow i\infty} = S_{AdS_5}+R^2\int d^2z\ \Bigg[ \sum_{i=1}^3 d\mu_i\wedge\star d\mu_i + \frac{\mu_1^2\mu_2^2\mu_3^2}{\alpha_1^2\mu_2^2\mu_3^2+\alpha_2^2\mu_1^2\mu_3^2+\alpha_3^2\mu_1^2\mu_2^2} d\omega\wedge\star d\omega \Bigg],
\end{equation}
where $\omega = \alpha_1\phi_1+\alpha_2\phi_2+\alpha_3\phi_3$.

\subsection{The pp-wave limit}

We shall now investigate the behavior of the theory in the limit of large $\gamma$ within the pp-wave limit. For this, we focus on the particular case where string concentrates at the equator of the sphere characterized by $\mu_1=1$. In this case, the metric of the sphere takes the form
\begin{equation}
g_{\gamma}\Big|_{\mu_1=1} = d\phi^2,
\end{equation}
which means we can  take the following curve as one geodesic of the strongly deformed $AdS_5\times S^5$ space:
\begin{equation}
u_5+iu_6 =  e^{iJ\tau} = \cos(J\tau) + i\sin (J\tau).
\end{equation}
We then proceed to examine perturbations around this geodesic. These perturbations can be described using the
parameterization
\begin{align}
u_{12} := u_1 +iu_2 = \nu e^{i\theta} =  \nu\cos(\theta) + i\nu\sin(\theta)\\
u_{34} := u_1 +iu_2 = \rho e^{i\psi} =  \rho\cos(\psi) + i\rho\sin(\psi).
\end{align}
The corresponding conserved currents are $j_{56} = J,\ \ j_{12} = \nu^2 d\theta$ and $j_{34} = \rho^2 d\psi$. 
Around this geodesic, the undeformed action in $AdS$ produces 4 massive modes, as shown in Appendix $\ref{penrose}$:
\begin{equation}\label{bmn}
S_b = R^2\frac{1}{4}\int d^2z \Str[J_2\wedge\star J_2] \rightarrow \int d^2z\ \partial \bar u_{12}\bar\partial u_{12} 
+ \partial \bar u_{34}\bar\partial u_{34} + J^2 (u_{12}^2+u_{34}^2).
\end{equation}
In other words, the action for the perturbation is described by a kinetic term for the complex fields $u_{12},u_{34}$, together 
with a mass term $J^2$. In terms of the coordinates $\nu,\rho,\theta,\psi$, the action $S_b$ is:
\begin{equation}
= \partial\nu\bar\partial\nu + \partial\rho\bar\partial\rho + \nu^2\partial\theta\bar\partial\theta + \rho^2\partial\psi\bar\partial\psi 
+ J^2(\nu^2+\rho^2).
\end{equation}
We now add the deformation. For the model (\ref{g1}), the correction is:
\begin{equation}
\delta L_b = -\frac{\gamma _1^2 \nu^2 \rho^2 \left(\nu^2 \partial\theta\bar\partial\theta+\rho^2\partial\psi\bar\partial\psi \right)}{1+\gamma_1^2 \nu^2 \rho^2} - \frac{\gamma_1\nu^2\rho^2}{1+\gamma_1^2 \nu^2 \rho^2}(\partial\theta\bar\partial\psi-\partial\psi\bar\partial\theta).
\end{equation}
The whole action is then
\begin{equation}
S_b^{\gamma_1} \rightarrow\int d^2z\left(\partial\nu\bar\partial\nu + \partial\rho\bar\partial\rho + \frac{\left(\nu^2\partial\theta\bar\partial\theta + \rho^2\partial\psi\bar\partial\psi -\gamma_1\nu^2\rho^2(\partial\theta\bar\partial\psi-\partial\psi\bar\partial\theta)\right)}{1+\gamma_1^2 \nu^2 \rho^2} + J^2(\nu^2+\rho^2)\right).
\end{equation}
Therefore, when we take the limit of large $\gamma_1$, we have:
\begin{equation}
S_b^{\gamma_1\rightarrow i\infty} \rightarrow\int d^2z\ \partial\nu\bar\partial\nu + \partial\rho\bar\partial\rho + J^2(\nu^2+\rho^2).
\hspace{2cm}
\end{equation}

This means that when we add our correction, deform the action, and take the large $\gamma$ correction, part of the 
action in equation (\ref{bmn}) disappears. Namely, the angular part given by $\nu^2 \partial\theta\bar\partial\theta+
\rho^2\partial\psi\bar\partial\psi$ vanishes in the action. This indicates some degeneracy of the metric in the large $\gamma$ 
limit.

Now, we turn on all deformation parameters, as in (\ref{beta}) and take the Penrose limit, obtaining
\begin{equation}
S_b^\gamma \rightarrow \frac{1}{4}\int d^2z\  \Bigg[(\partial\nu)^2 + (\partial\mu_3)^2 +J^2(\nu^2+\mu_3^2) \nonumber
\end{equation}
\begin{equation}
+ \frac{1}{4+4\gamma^2(\nu^2+\rho^2)}\Big( (\nu^2+\rho^2+\gamma^2\nu^2\rho^2)(\partial\phi_+)^2 + (\nu^2-\rho^2)(\partial\phi_+\bar\partial\phi_-+\partial\phi_-\bar\partial\phi_+) + (\nu^2+\rho^2)(\partial\phi_-)^2 \Big)\nonumber
\end{equation}
\begin{equation}
+ \frac{\gamma}{4+4\gamma^2(\nu^2+\rho^2)}\left(\nu^2\rho^2\ d\theta\wedge d\psi \right)\Bigg]\;,
\end{equation}
where $\phi_\pm = (\theta\pm\psi)$. The limit of large $\gamma$ provides:
\begin{equation}\label{penr3}
S_b^{\gamma\rightarrow i\infty} \rightarrow  \frac{1}{4}\int d^2\sigma\ \Bigg[ (\partial\nu)^2 + (\partial\rho)^2 + \frac{\nu^2\rho^2}{\nu^2+\rho^2}(\partial\phi_+)^2 +J^2(\nu^2+\rho^2)\Bigg].
\end{equation}
It is remarkable that taking the Penrose limit  in $AdS_5\times S^5$, we discover that there are eight massive string modes. 
The large $\gamma$ limit of the beta deformed string still has eight such massive modes, but now the action is degenerate at a 
given angle of propagation.

\subsection{Studying the matrix $\kappa_\gamma$}\label{diagonalization}

Let us analyze closely the deformed theory defined by the action (\ref{defaction}). Initially, the action is described by 
(\ref{S0}), and then the deformation is defined in (\ref{kappa}). This deformation involves the matrix $\kappa_\gamma=\left(B \frac{\gamma}{1-\gamma MB}\right)$, which plays a central role in the deformation. In this section, our focus will be on a careful examination of this matrix, aiming to understand the behavior of the deformed action as $\gamma$ becomes large. 
If the matrices $B$ and $M$ were invertible, the matrix $\kappa_\gamma$ at large $\gamma$ would simply reduce to $\kappa=-\frac{1}{M}$. However, this is not the case. The matrix $B$ employed in (\ref{Bm}) exhibits a non-trivial kernel characterized by $v_0=(1,1,1)$. We will then introduce an appropriate basis that facilitates a simpler representation of the matrix $\kappa_\gamma$.

The matrix $\kappa_\gamma$ involves two other ones, $B$ and $M$. 
First, we set the matrix $B$ to the form (\ref{Bm}) and matrix $M$  to (\ref{M}). We then define $K=M\cdot B$, 
\begin{equation}\label{MB}
K = M\cdot B =\left(
\begin{array}{ccc}
 M_{12,56}-M_{12,34} & M_{12,12}-M_{12,56} & M_{12,34}-M_{12,12} \\
 M_{34,56}-M_{34,34} & M_{34,12}-M_{34,56} & M_{34,34}-M_{34,12} \\
 M_{56,56}-M_{56,34} & M_{56,12}-M_{56,56} & M_{56,34}-M_{56,12} \\
\end{array}
\right)
\end{equation}
Given $K$, we can find a basis $\textbf{U}=\{v_0,v_+,v_-\}$ consisting of eigenvectors of $K$,
\begin{equation}\label{Kv}
K \cdot v_0 = 0\ ,\ \ \ K \cdot v_+ = \Gamma_+ v_+\ ,\ \ \ K \cdot v_- = \Gamma_- v_-.
\end{equation}
The eigenvectors are $v_0=(1,1,1)$ and $v_\pm$ are intricate expressions which are not necessary for the scope of this paper. The eigenvalues are given by $2\Gamma_\pm = \alpha\pm\beta$, with
\begin{align}\label{alphabeta}
&\alpha =-M_{12,34}-M_{34,56}-M_{56,12}+M_{56,34}+M_{34,12}+M_{12,56}\nonumber ,\\
&\beta^2 = \alpha{}^2+4 \Big( -M_{12,12} M_{34,34} -M_{12,12}M_{56,56} - M_{56,56}M_{34,34}\\
& +M_{12,12} \left(M_{34,56}+M_{56,34}\right)+M_{34,34}( M_{56,12} + M_{12,56})+M_{56,56}\left(M_{34,12} + M_{12,34}\right)\nonumber\\
&-M_{34,56} M_{56,12}-M_{34,12} M_{56,34}+M_{34,56} M_{56,34}-M_{12,56} M_{34,12}\nonumber\\
&+ M_{12,56}M_{56,12}-M_{12,56}M_{56,34}+M_{12,34}M_{34,12}-M_{12,34}M_{34,56}-M_{12,34}M_{56,12}\Big).\nonumber
\end{align}

For instance, when we restrict to the bosonic part, we have that $M_{12,12}=\mu_1^2$, $M_{34,34}=\mu_2^2$, $M_{56,56}=\mu_3^2$ and the remaining entries are zero, as computed in Appendix \ref{emb}. In this case we find that
\begin{equation}
\alpha = 0\ ,\hspace{1cm} \beta^2 = -4(\mu_1^2\mu_2^2+\mu_1^2\mu_3^2+\mu_2^2\mu_3^2).
\end{equation}

We now proceed to see how the matrix $\kappa_\gamma$ is written in terms of our new basis $\textbf{U}$. First, 
using (\ref{Kv}), we act with the matrix $\frac{1}{1-\gamma K}$ in this basis, obtaining 
\begin{align}
&\frac{1}{1-\gamma K}v_0 = v_0,\ \ \ \ \frac{1}{1-\gamma K}v_\pm = \frac{1}{1-\gamma\Gamma_\pm}v_\pm.
\end{align}
Then, the matrix $\kappa_\gamma = \gamma B \frac{1}{1-\gamma K}$ will act  in this basis as
\begin{align}\label{diagkappa}
&\kappa_\gamma\cdot v_0=0;\ \ \ \ \kappa_\gamma\cdot v_\pm = \frac{\gamma}{1-\gamma \Gamma_\pm}B\cdot v_\pm\,.
\end{align}
Finally, we conclude that $\kappa_\gamma$ in the basis $\textbf{U}$ takes the form
\begin{equation}\label{kappaU}
\big[\kappa_\gamma\big]_\textbf{U} = (v_-,Bv_+)\begin{pmatrix}
0&0&0\\
0&0&\frac{\gamma}{1-\gamma\Gamma_+}\\
0&-\frac{\gamma}{1-\gamma\Gamma_-}&0
\end{pmatrix}.
\end{equation}

Now, we write the deformation of the action (\ref{kappa}) using this language. Consider the current in terms of this new basis,
$j=j_0v_0+j_+v_++j_-v_-$, where $j_0,j_-,j_+$ are coefficients. We then use the expression in (\ref{kappaU}) for the matrix $
\kappa_\gamma$ to compute the product $\langle j,\kappa_\gamma\cdot j \rangle$. At the end, we obtain 
the deformation of the Lagrangian in the form
\begin{equation}\label{convergence}
\delta L =  \langle j,\kappa_\gamma\cdot j \rangle = j_+j_- \frac{\gamma^2\beta}{1 - \gamma\alpha+\gamma^2(\alpha^2-\beta^2)}
\langle v_+,Bv_-\rangle +  j_+j_- \frac{\gamma}{1 - \gamma\alpha+\gamma^2(\alpha^2-\beta^2)}\langle v_+,Bv_-\rangle\,.
\end{equation}
The advantage of expressing the Lagrangian in this manner is that one can demonstrate its convergence for large $\gamma$. 
To see that, we note that the possible divergence arises when $\alpha^2=\beta^2$, or equivalently, when $\alpha=\mp\beta$. In 
such cases, $\Gamma_\pm=0$, making the matrix $K$ degenerate along the $v_\pm$ directions. However, since the matrix 
$M$ is non-degenerate, the sole possibility is for $v_\pm$ to lie within the kernel of $B$. Yet, this cannot be the case, since $
\text{Ker}(B)$ possesses dimension one and already contains $v_0$, given by $v_0=(1,1,1)$.

Let us contrast our answer (\ref{kappaU}) to the initial guess at the beginning of this section, 
given by $-\frac{1}{M}$. In terms of the basis $\textbf{U}$, the latter is
\begin{equation}
-\left[\frac{1}{M}\right]_\textbf{U}=  \langle v_-,Bv_+\rangle\begin{pmatrix}
h_{11}&0&0\\
h_{12}&0&-\frac{1}{\Gamma_+}\\
h_{13}&\frac{1}{\Gamma_-}&0
\end{pmatrix},
\end{equation}
where $h_{11},h_{12},h_{13}$ are nonzero expressions, whose form is not useful to write here. As we commented, this is 
the answer when $B$ is invertible. The presence of a kernel in $B$ makes the $h$'s disappear.

\subsection{The BRST operator}\label{BRSTop}

In section 3, we studied how the BRST operator is modified under the $\gamma_i-$deformation. The goal of this section is 
to investigate the behavior of this operator under the regime of strong deformations. To accomplish this, we decompose the 
BRST operator into two distinct parts, as follows:
\begin{align}
&Q = \mathcal{Q}_0 + \gamma Q_1,\\
&\text{where}\ \ \ \mathcal{Q}_0 = Q_0+\hat{Q}_0,\ \text{with}\ \ Q_0 = (\lambda_1+\lambda_3)g\frac{\delta}{\delta g} -J_{1+}\frac{\delta}{\delta w_{1+}}-J_{3-}\frac{\delta}{\delta w_{3-}}\nonumber\\
&\hspace{5cm}\hat{Q}_0 = \left(\kappa^{ab}_\gamma j_{a+}(gt_bg^{-1})\right)_1\frac{\delta}{\delta w_{1+}} + \left(\kappa^{ab}_\gamma j_{a-}(gt_bg^{-1})\right)_3\frac{\delta}{\delta w_{3-}},\nonumber\\
&\ \ \ \text{and}\ \ \ Q_1 = B^{ab}\Lambda_at_b.\nonumber
\end{align}
The BRST operator is split into these two parts to accommodate different orders of the deformation parameter $\gamma$. 
Namely, $Q_1$ is linear in $\gamma$,
while $\mathcal{Q}_0$ is the zeroth order in $\gamma$,
\begin{equation}\label{Q0}
\frac{\mathcal{Q}_0}{\gamma}\xrightarrow[\gamma\rightarrow\infty]{}0.\ \ 
\end{equation}
This limit becomes apparent through the examination of the matrix $\kappa_\gamma$ for large values of $\gamma$, as illustrated in equation (\ref{kappaU}). We can now redefine the BRST charge as $q=\oint \frac{1}{\gamma}\mathcal{Q}_0 + Q_1$. Consequently, in the limit of large $\gamma$, the charge simplifies to $q_{f}=\oint Q_1$.

We started with an action and a BRST operator defined in the $AdS_5\times S^5$ space, and subsequently deformed both. 
Through our analysis in this section, we derived an explicit formula for the BRST operator at the large $\gamma$ limit.

The upcoming sections are devoted to investigating the action in the limit of large $\gamma$. In order to do that, we will use the fact 
that our theory remains BRST invariant in this limit, with the BRST charge given by $q_{f}=\oint Q_1$.

\subsection{Action for a deformation with strong coupling}

To study the full sigma model (\ref{def}) in the large $\gamma$ limit, we will use the symmetries of the action. First of all,
 we have a residual R-symmetry corresponding to rotations under the planes $12$, $34$ and $56$, defined in (\ref{coord}). Restricted to the bosonic part, these are shifts of the angles, $\frac{\partial}{\partial\phi_1},\frac{\partial}{\partial\phi_2},\frac{\partial}{\partial\phi_3}$. Therefore, since the bosonic action is invariant under the rotation of these planes, the angle variables $\phi_i$ only enter through their derivatives. In other words, the only terms in the bosonic part of the action describing the angular part are kinetic terms. Moreover, we also see that some of these kinetic terms disappear (see equations (\ref{f1}) and (\ref{f123})). We will explain why this happens based on the BRST symmetry of the action we just described in the last section. Furthermore, based on the BRST structure we will prove a statement concerning the degeneracy of the sigma model background field.

First, we will formulate a general statement for pure spinor theories only based on its BRST operator.

\subsection{BRST structure and the action for more general backgrounds}

We are studying a theory with a BRST operator given by $Q_1=B^{ab}\Lambda_at_b$, as shown above in section (\ref{BRSTop}). However, we don't have an explicit expression for the action. In this section, we will study some properties of this action only using its invariance under the BRST operator $Q_1$.

We will start with a pure spinor sigma model in a particular background that allows us to integrate out the conjugate momenta. Separating the terms with ghosts and without ghosts in the action, this model can be writen as:
\begin{equation}\label{Sf}
S = \int d^2z\ A_{\mu\nu}\partial_+Z^\mu\partial_-Z^\nu + S_g(\lambda,w),
\end{equation}
where $A_{\mu\nu}(Z)=G_{\mu\nu}(Z)+B_{\mu\nu}(Z)$ and $Z^\mu=(Z^m,\theta^\alpha,\theta^{\dot\alpha})$ 
are local coordinates of the superspace. 

The AdS sigma model and its deformations fall under this category. The case of interest in this paper arises from a 
deformation of AdS in a specific limit of strong deformation. Therefore, our conclusions in this section apply to the 
case of interest.

Back to the model (\ref{Sf}), we will use the BRST symmetry inherent in the pure spinor action \cite{BH} to extract information regarding the constraints that the background fields must satisfy. The BRST operator can be expressed as
\begin{equation}
Q = Q^\mu\frac{\partial}{\partial Z^\mu} + Q_\lambda^\alpha\frac{\partial}{\partial\lambda^\alpha} +  Q_w^\alpha\frac{\partial}{\partial w^\alpha}.
\end{equation}
Since this operator has ghost number $1$, the matter part can be writen as:
\begin{equation}
Q^\mu\frac{\partial}{\partial Z^\mu} = \lambda^\alpha E_\alpha^\mu\frac{\partial}{\partial Z^\mu},
\end{equation}
where $E_\alpha$ is a function of space-time coordinates $Z$. Let us also split the BRST operator into left and right sectors,
\begin{equation}
Q = Q_L + Q_R,\ \ \ Q^\mu\frac{\partial}{\partial Z^\mu} = \lambda_L^{\alpha} E_{L\alpha}^\mu\frac{\partial}{\partial Z^\mu}+\lambda_R^{\alpha} E_{R\alpha}^\mu\frac{\partial}{\partial Z^\mu}.
\end{equation}
The BRST variation of $w$ and $\lambda$ in $S_g$ will possibly cancel with the BRST variation of the matter part. However, it can be argued that $S_g$ is BRST invariant when the pure spinors $\lambda$ and their conjugate momentum $w$ are on-shell. To demonstrate this, we use the equations of motion for the pure spinors $\lambda_L$ and $\lambda_R$, which are determined by the variation of $w$: $\frac{\delta}{\delta w^\alpha}S_g=0$. Also, we use the equations of motion for the conjugate momentum $w_+$ and $w_-$, given by the variation of $\lambda$, $\frac{\delta}{\delta \lambda^\alpha}S_g=0$. Therefore, we have
\begin{equation}
\left(Q_w^\alpha\frac{\delta}{\delta w^\alpha}+Q_\lambda^\alpha\frac{\delta}{\delta \lambda^\alpha}\right)S_g=0,\ \ \ \ \ \text{for $\lambda$ and $w$ on-shell}.
\end{equation}
The BRST variation of the matter fields $Z^\mu$ in $S_g$ will result in an expression containing $w$-ghost fields. This variation cannot cancel out with the BRST variation of the matter part of the action because the latter does not involve $w$-ghosts. Therefore it vanishes on its own:
\begin{equation}\label{QzSg}
Q^\mu\frac{\delta}{\delta Z^\mu}S_g=0,\ \ \ \ \ \text{for $\lambda$ and $w$ on-shell}.
\end{equation}
We conclude then that the complete BRST variation of the ghost action $S_g$ is zero on-shell,
\begin{equation}
QS_g = 0,\ \ \ \ \ \text{for $\lambda$ and $w$ on-shell}.
\end{equation}
Therefore, from $QS=0$, we also have
\begin{align}\label{QA=0}
&Q\int d^2\tau A_{\mu\nu}\partial_+Z^\mu\partial_-Z^\nu = 0,\ \ \ \ \text{for } \lambda\ \text{and } w\ \text{on-shell}.
\end{align}
Expanding this equation, we obtain
\begin{align}\label{QA2}
&Q\int d^2\tau A_{\mu\nu}\partial_+Z^\mu\partial_-Z^\nu = \int d^2\tau\Big( Q^\rho\partial_\rho A_{\mu\nu}\partial_+Z^\mu\partial_-Z^\nu + A_{\mu\nu}\partial_+Q^\mu\partial_-Z^\nu +A_{\mu\nu}\partial_+Z^\mu\partial_-Q^\nu\Big)\nonumber\\
&\hspace{1cm}=\int d^2\tau\Big( Q^\rho\partial_\rho A_{\mu\nu}\partial_+Z^\mu\partial_-Z^\nu + A_{\mu\nu}\lambda^\alpha\partial_+E_\alpha^\mu\partial_-Z^\nu  +A_{\mu\nu}\lambda^\alpha\partial_+Z^\mu\partial_-E_\alpha^\nu\Big)\\
&\hspace{4cm} + A_{\mu\nu}E^\mu_\alpha\partial_+\lambda^\alpha\partial_-Z^\nu  +A_{\mu\nu}E^\nu_\alpha\partial_+Z^\mu\partial_-\lambda^\alpha\Big)=0,\hspace{1cm}\text{for } \lambda\ \text{and } w\ \text{on-shell}\nonumber
\end{align}

This equality is true for any field configuration of $Z$ and any on-shell field configuration for $\lambda$ and $w$. First we consider a field $Z^\mu$ which does not depend on $\tau^-$, with $\partial_-Z=0$. In this case, equation (\ref{QA2}) reduces to
\begin{align}\label{d-Z=0}
\int d^2\tau A_{\mu\nu}E^\nu_\alpha\partial_+Z^\mu\partial_-\lambda^\alpha=0,\ \ \ \ \ &\text{for $\lambda$ on-shell},\\
&\text{and $Z$ such that }\partial_-Z=0.\nonumber
\end{align}
We further specify the field configuration of $Z^\mu$ by setting it to:
\begin{equation}\label{Zconf}
Z^\mu = Z^\mu_0+\eta V^\mu H(\tau^+),
\end{equation}
where $V^\mu$ is an arbitrary constant vector in the superspace, $\eta$ is a small parameter and $H(x)\equiv 1_{x\ge0}$ is the Heaviside step function. The purpose of using the Heaviside function is that its derivative is a delta function, $\partial_+H(\tau^+)=\delta(\tau^+)$. This will be used to eliminate the integral in the equation (\ref{d-Z=0}).

\begin{figure}
    \centering

\tikzset{every picture/.style={line width=0.75pt}} %set default line width to 0.75pt        

\begin{tikzpicture}[x=0.75pt,y=0.75pt,yscale=-1,xscale=1]
%uncomment if require: \path (0,393); %set diagram left start at 0, and has height of 393

%Curve Lines [id:da5021317260456265] 
\draw    (148,203) .. controls (203,192) and (222,23) .. (315,26) ;
%Curve Lines [id:da8189063745541679] 
\draw    (269,266) .. controls (324,255) and (372,79) .. (431,74) ;
%Curve Lines [id:da21779320376457323] 
\draw    (148,203) .. controls (172,228) and (232,226) .. (269,266) ;
%Curve Lines [id:da22554048964481188] 
\draw    (315,26) .. controls (336,51) and (409,40) .. (431,74) ;
%Curve Lines [id:da8340941254587053] 
\draw [color={rgb, 255:red, 14; green, 13; blue, 13 }  ,draw opacity=1 ]   (170,200) .. controls (188.62,222.54) and (257.18,226.83) .. (277.8,251.47) ;
\draw [shift={(279,253)}, rotate = 233.84] [color={rgb, 255:red, 14; green, 13; blue, 13 }  ,draw opacity=1 ][line width=0.75]    (10.93,-3.29) .. controls (6.95,-1.4) and (3.31,-0.3) .. (0,0) .. controls (3.31,0.3) and (6.95,1.4) .. (10.93,3.29)   ;
%Curve Lines [id:da49036835553287406] 
\draw [color={rgb, 255:red, 7; green, 7; blue, 7 }  ,draw opacity=1 ]   (170,200) .. controls (193.76,193.07) and (232.22,73.43) .. (289.27,39.99) ;
\draw [shift={(291,39)}, rotate = 151.11] [color={rgb, 255:red, 7; green, 7; blue, 7 }  ,draw opacity=1 ][line width=0.75]    (10.93,-3.29) .. controls (6.95,-1.4) and (3.31,-0.3) .. (0,0) .. controls (3.31,0.3) and (6.95,1.4) .. (10.93,3.29)   ;
%Curve Lines [id:da37852561126546946] 
\draw [color={rgb, 255:red, 23; green, 39; blue, 207 }  ,draw opacity=1 ]   (213,117) .. controls (265,95) and (274,160) .. (345,173) ;
%Curve Lines [id:da9194359615940172] 
\draw [color={rgb, 255:red, 13; green, 230; blue, 33 }  ,draw opacity=1 ]   (260.96,210.94) .. controls (300.96,180.94) and (308.96,107.94) .. (348.96,77.94) ;
%Curve Lines [id:da27070309140971527] 
\draw [color={rgb, 255:red, 13; green, 230; blue, 33 }  ,draw opacity=1 ]   (207.96,187.94) .. controls (247.96,157.94) and (255.96,84.94) .. (295.96,54.94) ;
%Curve Lines [id:da22548538754725767] 
\draw [color={rgb, 255:red, 13; green, 230; blue, 33 }  ,draw opacity=1 ]   (234.96,198.94) .. controls (274.96,168.94) and (282.96,95.94) .. (322.96,65.94) ;
%Curve Lines [id:da258719012458008] 
\draw [color={rgb, 255:red, 13; green, 230; blue, 33 }  ,draw opacity=1 ]   (284.96,226.94) .. controls (324.96,196.94) and (332.96,123.94) .. (372.96,93.94) ;
%Curve Lines [id:da26027806608979565] 
\draw    (310,158) .. controls (327.82,120.38) and (395.03,114.88) .. (442.95,129.33) ;
\draw [shift={(444.4,129.77)}, rotate = 197.35] [color={rgb, 255:red, 0; green, 0; blue, 0 }  ][line width=0.75]    (10.93,-3.29) .. controls (6.95,-1.4) and (3.31,-0.3) .. (0,0) .. controls (3.31,0.3) and (6.95,1.4) .. (10.93,3.29)   ;
%Curve Lines [id:da1417457494758252] 
\draw [color={rgb, 255:red, 22; green, 230; blue, 11 }  ,draw opacity=1 ]   (301,215) .. controls (313.81,234.7) and (354.75,258.28) .. (391.33,240.83) ;
\draw [shift={(393,240)}, rotate = 152.82] [color={rgb, 255:red, 22; green, 230; blue, 11 }  ,draw opacity=1 ][line width=0.75]    (10.93,-3.29) .. controls (6.95,-1.4) and (3.31,-0.3) .. (0,0) .. controls (3.31,0.3) and (6.95,1.4) .. (10.93,3.29);

% Text Node
\draw (296,29.4) node [anchor=north west][inner sep=0.75pt]  [color={rgb, 255:red, 10; green, 10; blue, 10 }  ,opacity=1 ]  {$\textcolor[rgb]{0.05,0.04,0.04}{\tau ^{+}}$};
% Text Node
\draw (276,227.4) node [anchor=north west][inner sep=0.75pt]  [color={rgb, 255:red, 14; green, 12; blue, 12 }  ,opacity=1 ]  {$\textcolor[rgb]{0.02,0.02,0.02}{\tau ^{-}}$};
% Text Node
\draw (354.11,163.89) node [anchor=north west][inner sep=0.75pt]  [rotate=-0.72]  {$\textcolor[rgb]{0.02,0.38,0.97}{S\ }$};
% Text Node
\draw (450.4,124.17) node [anchor=north west][inner sep=0.75pt]    {$\lambda _{1}^{\alpha }\Bigl|_{S} =\ u^{\alpha } H \left( \tau ^{-}\right)$};
% Text Node
\draw (372,54.4) node [anchor=north west][inner sep=0.75pt]    {$\Sigma $};
% Text Node
\draw (401,231.4) node [anchor=north west][inner sep=0.75pt]    {$\textcolor[rgb]{0.15,0.83,0.09}{L_{y}}$};
\end{tikzpicture}
    \caption{Worldsheet $\Sigma$ parametrized by $\tau^+$ and $\tau^-$. Cauchy surface $S$ defined on the worldsheet and the initial boundary condition for the fields $\lambda_1^\alpha(\tau^+,\tau^-)$. Green lines represent those transverse to the Cauchy surface $S$ along $\Sigma$.}
    \label{Worldsheet}
\end{figure}
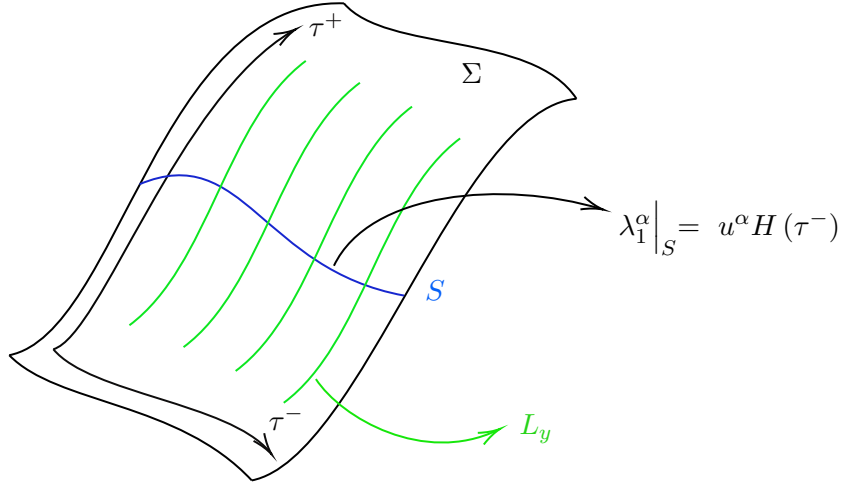

Let us now deal with the pure spinor fields, which are on-shell, so we have to be careful if we want to specify a field 
configuration for them. First, we choose $\lambda_3=0$, since this is one possible solution to the equation of motions of 
$\lambda_3$. Then, we notice that $w_+$ has conformal weight $(1,0)$ and $w_-$ has conformal weight $(0,1)$. Therefore, roughly speaking, the action of conformal weight $(1,1)$ and ghost number zero contains only terms of the form $(w_-\lambda_1)\cdots$, $(w_-\partial_+\lambda_1)\cdots$, $(w_+\lambda_3)\cdots$, $(w_+\partial_-\lambda_3)\cdots$ and $(w_-\lambda_1)(w_+\lambda_3)\cdots$. Then, the equation of motion for $\lambda_1$ becomes
\begin{equation}\label{difeq}
\partial_+\lambda^\alpha_1=\lambda_1^\beta E_{\beta\mu}^\alpha(Z)\partial_+Z^\mu,
\end{equation}
where $E_{\beta\mu}^\alpha$ are complicated functions of $Z$, obtained from varying the action with respect to $w_{3-}$ and setting $\lambda_3=0$. This is a partial differential equation and we will now proceed to formulate a Cauchy problem with it. Consider the differential equation (\ref{difeq}) and a Cauchy surface $S$ defined as a one dimensional surface in the worldsheet $\Sigma$ by
\begin{equation}
S \equiv \big\{ (\tau^+,\tau^-);\ \ \tau^+=0 \big\}\subset\Sigma.
\end{equation}
Then, we define the boundary condition of the problem by defining the functions $\lambda_1^\alpha$ 
on the Cauchy surface $S$,
\begin{equation}
\lambda^\alpha_1(\tau^+=0,\tau^-) \equiv u^\alpha H(\tau^-),
\end{equation}
where $u^\alpha$ is a constant pure spinor. This is schematically represented in figure \ref{Worldsheet}.

Now, we have fully stated our initial value problem. In summary:
\begin{align}\label{cauchy}
&\partial_+\lambda^\alpha_1=\lambda_1^\beta E_{\beta\mu}^\alpha(Z)\partial_+Z^\mu,\nonumber\\
&\lambda_1^\alpha\Big|_{S=\{\tau^+=0\}} = u^\alpha H(\tau^-).
\end{align}
Outside the Cauchy surface, the functions $\lambda_1^\alpha$ will be determined by the differential equation. To do that, 
we solve the differential equation along each green line in figure \ref{Worldsheet}, defined by
\begin{equation}
L_y \equiv \{ (\tau^+,\tau^-)\ ;\ \tau^-=y\}\ ,\ \ \ \forall\ (0,y)\in S
\end{equation}
First, we remember that we have specified the field configuration for $Z$ in equation (\ref{Zconf}). We rewrite the 
differential equation with this field configuration,
\begin{equation}\label{diff}
\partial_+\lambda^\alpha_1(\tau^+,\tau^-)=\eta\lambda_1^\beta(\tau^+,\tau^-) E_{\beta}^\alpha(V)\delta(\tau^+),
\end{equation}
where we denote $E_\alpha^\beta=V^\mu E_{\alpha\mu}^\beta$. Now, for each line $L_{y}$ we have an ordinary differential equation, with initial value at the point $L_y\cap S$. Along the $L_y$ lines, we define the functions $\lambda_{1y}^\alpha(\tau^+) = \lambda^\alpha(\tau^+,y)$, and ordinary differential equations for them:
\begin{align}
&\frac{d}{d\tau^+}\lambda_{1y}^\alpha(\tau^+) = \eta\lambda_{1y}^\beta(\tau^+)E_\beta^\alpha\delta(\tau^+),\\
&\lambda_{1y}^\alpha(0) = u^\alpha H(y).\nonumber
\end{align}
The solution for a fixed $y$ is
\begin{equation}
\lambda_{1y}^\alpha(\tau^+) = u^\beta H(y)\Big[e^{\eta E\int_0^{\tau^+}\delta(x)dx}\Big]^\alpha_\beta.
\end{equation}
Indeed, we can check that it is a solution, since 
\begin{equation}
\frac{d}{d\tau^+}\lambda^\alpha_{1y}(\tau^+) = u^\beta H(y) \Big[e^{\eta E\int_0^{\tau^+}\delta(x)dx}\Big]^\gamma_\beta \eta E^\alpha_\gamma \delta(\tau^+) = \eta\lambda^\beta_{1y}(\tau^+)E_\beta^\alpha\delta(\tau^+),
\end{equation}
and so it satisfies the boundary conditions $\lambda^\alpha_{1y}(0) = u^\alpha H(y)$. Now, we can vary the 
parameter $y$ and write the full solution as
\begin{equation}\label{lambda1}
\lambda_{1}^\alpha(\tau^+,\tau^-) = u^\beta H(\tau^-)\Big[e^{\eta E\int_0^{\tau^+}\delta(x)dx}\Big]^\alpha_\beta.
\end{equation}

Now that we have specified the field configurations for $Z$ and $\lambda$, we go back to the expression of the BRST variation 
(\ref{d-Z=0}). Explicitly, we obtain for the derivatives 
\begin{equation}
\partial_+Z^\mu\partial_-\lambda_1^\alpha = \eta V^\mu u^\alpha \delta(\tau^+)\delta(\tau^-)\;,
\end{equation}
where we consider $\eta^2=0$. After integration, (\ref{d-Z=0}) reduces to
\begin{equation}
\Big(A_{\mu\nu}E^\nu_{R\alpha}V^\mu u^\alpha\Big)\Big|_{Z=Z_0}=0.
\end{equation}

If we instead specify the field configurations to be $Z^\mu=Z_0^\mu+\eta V^\mu H(\tau^-)$, $\lambda^\alpha_1=0$ and a 
non-vanishing on-shell spinor $\lambda_3$, corresponding to (\ref{lambda1}), 
\begin{equation}\label{lambda3}
\lambda_{3}^\alpha(\tau^+,\tau^-) = u^\beta H(\tau^+)\Big[e^{\eta E\int_0^{\tau^-}\delta(x)dx}\Big]^\alpha_\beta,
\end{equation}
we get a new corresponding condition, that is, equation (\ref{QA2}) reduces to
\begin{equation}
\Big(A_{\mu\nu}E^\mu_{L\alpha} V^\nu u^\alpha\Big)\Big|_{Z=Z_0}=0.
\end{equation}
The vector $V^\mu$ is arbitrary and we can take it out of the equation. In summary, we have
\begin{equation}\label{AdL}
A_{\mu\nu}E^\mu_{L}\langle u\rangle = 0,
\end{equation}
\begin{equation}\label{AdR}
A_{\mu\nu}E^\nu_{R}\langle u\rangle = 0,
\end{equation}
where angle brackets $\langle u\rangle$ means a linear dependence of the operator on $u$. Although it is written in terms of a 
set of coordinates, this result is independent of the coordinate system we use, owing to the invariance of equations (\ref{AdL}) 
and (\ref{AdR}) under coordinate transformations. To see this, we notice that under a change of coordinates $Z\mapsto Z'$, $A_{\mu\nu}$ transforms as a 2-tensor and $E^\mu$ transforms as a vector. Now, applying a change of coordinates to the equation $A_{\mu\nu}E^\nu=0$ yields
\begin{equation}
0=\frac{\partial Z^\mu}{\partial Z'^{\mu'}}A_{\mu\nu}E^\nu = \frac{\partial Z^\mu}{\partial Z'^{\mu'}}A_{\mu\rho}\frac{\partial Z^\rho}{\partial Z'^{\nu'}}\frac{\partial Z'^{\nu'}}{\partial Z^{\sigma}}E^\sigma = A_{\mu'\nu'}E^{\nu'}.
\end{equation}

When $u$ runs over the pure spinor cone, $E_{L,R}^\mu\langle u\rangle$ spans a subset in $T(AdS_5\times S^5)$, which we call
\begin{align}
&\mathcal{Z}_{L,R} = \{ d_{L,R}^\mu\langle u\rangle, \ \ u\in\text{PS-cone} \}\;,
\end{align}
and since $A_{\mu\nu}=G_{\mu\nu}+B_{\mu\nu}$, we have in general
\begin{equation}
\mathcal{Z}_{L}\in \ker(G+B)
\end{equation}
\begin{equation}
\mathcal{Z}_{R}\in \ker(G-B)
\end{equation}
It should also be possible to derive this result from the generalized geometry formalism in \cite{zavaleta2019pure}.

\subsection{Sigma model under strong coupling deformation: large $\gamma$ limit}

Now we are able to study the fishnet limit properly by applying what we have learned from the generic case in the last section. Consider our model applied to the Maldacena-Lunin-Frolov case. The BRST operator of our model in the strong deformation regime is given by

\begin{equation}
Q_1 = B^{[ij][kl]}\Lambda_{ij}t_{kl},
\end{equation}
where $B^{[ij][kl]}$ is a constant matrix given by (\ref{gammaB}).
To simplify, let us examine the simplest case, where only one parameter, $\gamma_3$,  is non-vanishing, while $\gamma_1=\gamma_2=0$, as illustrated in (\ref{g1}). In this case, the BRST operator is
\begin{equation}\label{gamma1brst}
Q_1 = \Lambda_{12}t_{34} - \Lambda_{34}t_{12}.
\end{equation}

We have an explicit expression for the generators of rotations using the embedding bosonic coordinates in (\ref{tij}). When we include fermionic variables we have corrections: $t_{ij}=u_{[i}\frac{\partial}{\partial u^{j]}} + \text{[fermions]}$. Then, the first terms in the BRST operator are
\begin{equation}\label{bosons}
Q_1= \Lambda_{12}\frac{\partial}{\partial \phi_2} - \Lambda_{34}\frac{\partial}{\partial \phi_1} + \cdots
\end{equation}

In general, we write $\Lambda_a = \Str[(g^{-1}t_\alpha g)t_a]\lambda^\alpha=(g^{-1}t_\alpha g)_a\lambda^\alpha$, where we denote $t_\alpha$ as the generators of $\mathfrak{g}_1\oplus\mathfrak{g}_3$ and $\lambda = \lambda_3-\lambda_1$. Then, the BRST operator can be written as
\begin{equation}
Q_1 = \lambda^\alpha(g^{-1}t_\alpha g)_{12}t^\mu_{34}\frac{\partial}{\partial Z^\mu} - \lambda^\alpha(g^{-1}t_\alpha g)_{34}t^\mu_{12}\frac{\partial}{\partial Z^\mu} 
\end{equation}

From this, we see that the two spaces $\mathcal{Z}_L$ and $\mathcal{Z}_R$ are actually the same space, explicitly given by
\begin{equation}
\mathcal{Z} = \{ (g^{-1}ug)_{12}t^\mu_{34} - (g^{-1}ug)_{34}t^\mu_{12}, \ \ u\in\text{PS-cone} \}
\end{equation}
Then, the conditions (\ref{AdL}) and (\ref{AdR}) are written as
\begin{equation}\label{alpha}
A_{\mu\nu}t_{34}^\mu (g^{-1}u g)_{12}-A_{\mu\nu}t_{12}^\mu (g^{-1}u g)_{34}=0
\end{equation}
\begin{equation}\label{alpha2}
A_{\mu\nu}t_{34}^\nu (g^{-1}u g)_{12}-A_{\mu\nu}t_{12}^\nu (g^{-1}u g)_{34}=0.
\end{equation}

The reason why the spaces $\mathcal{Z}_L$ and $\mathcal{Z}_R$ are the same is because they are both proportional to the same $B^{ab}t_b$. Indeed, we now prove that this space is generated by the vectors $B^{ab}t_b$. To see that, we define a surjective map that acts by
projecting $\Ad_g(u)$ in the 2-dimensional subspace generated by $t_{12}$ and $t_{34}$, with $u$ a constant pure spinor,
\begin{align}\label{surjmap}
&\text{PS-cone}\subset\mathbb{R}^{16}\longrightarrow\mathbb{R}^2\\
&\hspace{2cm}u\longmapsto \sum_{a=12,34}(g^{-1}ug)_at_a.\nonumber
\end{align}

We can then find an element $u_0$ such that the image of it under this map is in the subspace of $\mathbb{R}^2$ generated by $t_{34}$. We can always find such an element because this map is surjective. This is explained in Appendix (\ref{B1}). Therefore, the condition (\ref{alpha}) reduces to
\begin{equation}
A_{\mu\nu}t_{12}^\mu u_0^\alpha(g^{-1}t_\alpha g)_{34}=0,
\end{equation}
which then implies that $A_{\mu\nu}t_{34}^\mu u_0^\alpha(g^{-1}t_\alpha g)_{12}=0$. Therefore, we get that the 
initial constraints from BRST invariance of the action simplify to the following constraint for $A_{\mu\nu}$:
\begin{equation}\label{isometry1}
A_{\mu\nu}t_{12}^\mu=A_{\mu\nu}t_{34}^\mu=0.
\end{equation}
If we do the same procedure for equation (\ref{alpha2}), we obtain a corresponding condition,

\begin{equation}\label{isometry2}
A_{\mu\nu}t_{12}^\nu=A_{\mu\nu}t_{34}^\nu=0
\end{equation}
These equations are telling us that $A_{\mu\nu}$ is degenerate in the $t_{34}$ and $t_{56}$ directions. In the two 
equations above, we can take the symmetric and the antisymmetric parts. Doing that, we translate the two conditions in 
$A$ into two conditions for the metric $G$, and another two for the $B$-field,
\begin{equation}\label{isometry}
G_{\mu\nu}t_{12}^\mu=G_{\mu\nu}t_{34}^\mu=0,\ \ \ B_{\mu\nu}t_{12}^\mu=B_{\mu\nu}t_{34}^\mu=0.
\end{equation}

The field $A$ exhibits a degeneracy, which is already manifested in the bosonic expression (\ref{f1}). In the bosonic case, 
this degeneracy appears in the angular component of the metric. This arises because the action of the BRST operator on 
the bosonic part is proportional to shifts in the angles, as indicated in (\ref{bosons}). For the entire action, a broader statement 
is explained by equations (\ref{isometry}). This degeneracy corresponds, reciprocally, to the selective decoupling of certain 
fields, as discussed in \cite{Gurdogan:2015csr}. Specifically, equation (\ref{isometry}) dictates that the structure 
of the action is of the type
\begin{equation}
S = \int d^2z\ A_{\phi_3\phi_3}(Z)\partial\phi_3\bar\partial\phi_3 + \xCcancel[red]{\sum_{i,j=1}^2A_{\phi_i\phi_j}(Z)\partial\phi_i\bar\partial\phi_j} + \cdots 
\end{equation}

\subsubsection{General picture of degeneracy and relation to Maldacena-Lunin-Frolov}

In order to understand better what is happening to the background fields in the fishnet limit, we will formulate a general 
picture for it. This will be useful in order to understand the theory when we turn on all the three deformation parameters $
\gamma_1,\gamma_2,\gamma_3$. We will also see how this general picture explains the structure of the 
Maldacena-Lunin-Frolov solution in the large $\gamma$ limit, as written in equation (\ref{f123}). 

We begin by applying equations (\ref{AdL}) and (\ref{AdR}) to the sigma model with three deformation parameters with a strong coupling, namely

\begin{equation}
A_{\mu\nu}T^\mu\langle u\rangle = A_{\mu\nu}T^\nu\langle u\rangle=0,
\end{equation}
where $T^\mu$ is a space-time vector field which depends linearly to the parameter $u$, defined by $T^\mu\langle 
u\rangle=t^\mu_aB^{ab}(g^{-1}t_\alpha g)_bu^\alpha$. As argued in the last section, the spaces $\mathcal{Z}_L$ and $
\mathcal{Z}_R$ are spanned by $B^{ab}t_b$. In turn, when $u$ runs over the pure spinor cone, $T^\mu$ spans the  subspace 
of $T(AdS_5\times S^5)$
\begin{equation}\label{Z123}
\mathcal{Z} = \{ t^\mu_a W^a\ | \ \ \text{for }\  W \in \Im(B^{})  \},
\end{equation}
where $\Im(B^{})$ means the Image of the matrix $B^{ab}$. In our case, $W$ is any vector of the form $W^a=B^{ab}C_b$. $\mathcal{Z}$ is the space of degeneracy of the background fields.

Next, we write the three parameters as $\gamma_1=\alpha_1\gamma$, $\gamma_2=\alpha_2\gamma$ and 
$\gamma_3=\alpha_3\gamma$, where $\alpha_i$ are non-zero constants and $\gamma$ is the parameter 
we will tend to infinity. The BRST operator is then
\begin{align}
Q = & \left( \alpha_2\Lambda_{56} - \alpha_3\Lambda_{34} \right)t_{12}\nonumber\\
+& \left( \alpha_3\Lambda_{12} - \alpha_1\Lambda_{56} \right)t_{34}\nonumber\\
+& \left( \alpha_1\Lambda_{34} - \alpha_2\Lambda_{12} \right)t_{56}.
\end{align}
If we redefine the generators $t_a$ by $\mathcal{T}_1 := t_{12}$, $\mathcal{T}_2 = t_{34}-\frac{\alpha_2}{\alpha_1}t_{12}$ and $\mathcal{T}_3 := t_{56}-\frac{\alpha_3}{\alpha_2}t_{34}$, we will have
\begin{align}\label{angleredefinition}
&t_{12} = \mathcal{T}_1,\nonumber\\
&t_{34} = \frac{\alpha_2}{\alpha_1}\mathcal{T}_1+\mathcal{T}_2,\\
&t_{56} = \frac{\alpha_3}{\alpha_1}\mathcal{T}_1+\frac{\alpha_3}{\alpha_2}\mathcal{T}_2+\mathcal{T}_3,\nonumber
\end{align}
and therefore the BRST operator can be written as
\begin{equation}\label{brst3}
Q = \frac{\alpha_1}{\alpha_2}(\alpha_3\Lambda_{34}-\alpha_2\Lambda_{56})\mathcal{T}_2 
+ (\alpha_1\Lambda_{34}-\alpha_2\Lambda_{12})\mathcal{T}_3
\end{equation}
In the action (\ref{Sfrolov}), the remaining angular term in the large $\gamma$ limit is $d\omega^2$, where 
$\omega = \sum_{i=1}^3\alpha_i\phi_i$, as explicitly written in (\ref{f123}),
\begin{equation}
R^2\int d^2z\ \Bigg[ \sum_{i=1}^3 d\mu_i\wedge\star d\mu_i + \frac{\mu_1^2\mu_2^2\mu_3^2}{\alpha_1^2\mu_2^2\mu_3^2+\alpha_2^2\mu_1^2\mu_3^2+\alpha_3^2\mu_1^2\mu_2^2} d\omega\wedge\star d\omega \Bigg],
\end{equation}
Then, the degeneracy space $\mathcal{Z}$ is generated by linear combinations of $t_{12},t_{34},t_{56}$ on which the 1-form $d\omega = \sum_{i=1}^3\alpha_id\phi_i$ is zero. These combinations is precisely given by the vectors appearing in the BRST operator (\ref{brst3}),
\begin{equation}
d\omega(\mathcal{T}_2)=d\omega(\mathcal{T}_3)=0.
\end{equation}
An interesting observation is that the 1-form $d\omega$ is in the kernel of $B$. Indeed, the fact that $B$ has a kernel of 
dimension 1 implies that the space of degeneracy defined in (\ref{Z123}) has only 2 dimensions. As a consequence of this fact, 
some kinetic terms of the angles still survive, for instance in the bosonic case, where precisely $d\omega^2$ survives.

In summary, in the general case the kinetic term for the angle of the $12$ plane will not vanish in the bosonic action because its 
generator is no longer present in the BRST operator, as evidenced in (\ref{brst3}). On the other hand, the planes $\mathcal{T}
_2$ and $\mathcal{T}_3$ define angles that have vanishing kinetic terms. The angle of the $12$ plane is not a special 
direction, but only a consequence of the coordinates redefinition $(\ref{angleredefinition})$. In fact, we could have chosen any 
new coordinate redefinition such that some other angle remains in the action. This can be seen in the Penrose limit of this 
case, (\ref{penr3}). The general condition for the field $A_{\mu\nu}$ now is
\begin{equation}\label{TA=01}
A_{\mu\nu}\mathcal{T}^\mu_2=A_{\mu\nu}\mathcal{T}^\mu_3=0
\end{equation}
\begin{equation}\label{TA=02}
A_{\mu\nu}\mathcal{T}^\nu_2=A_{\mu\nu}\mathcal{T}^\nu_3=0,
\end{equation}
or
\begin{equation}\label{deg}
G_{\mu\nu}\mathcal{T}^\mu_2=G_{\mu\nu}\mathcal{T}^\mu_3=0,\ \ \ \ B_{\mu\nu}\mathcal{T}^\mu_2=B_{\mu\nu}\mathcal{T}^\mu_3=0.
\end{equation}

Then, in the fishnet model $d_{L}^\mu\langle u\rangle=d_{R}^\mu\langle u\rangle=t^\mu_aB^{ab}(g^{-1}ug)_b$. In this case, $\mathcal{Z}_L$ and $\mathcal{Z}_R$ collapses to the same space $\mathcal{Z}$, which by itself has the interpretation of a degeneration space of the background fields.

\subsubsection{Relation of degeneracy with BRST variation of $w$}
In the last subsection, we supposed that we don't know the structure of the BRST operator $Q_1$ acting on the 
$w$-ghost. In fact, the degeneracy of the background fields will imply that BRST acting on $w$ is zero. Let us prove it.

Without assuming the structure of BRST transformation of $w$, the background fields satisfy 
the relations (\ref{TA=01}) and (\ref{TA=02}),
\begin{equation}\label{TA=0}
A_{\mu\nu}\mathcal{T}^\mu_{a}=A_{\mu\nu}\mathcal{T}^\nu_{a}=0\;,
\end{equation}
for $a=2,3$. Now, this equation implies that the matter part of the action is BRST invariant off-shell. This happens 
because the degeneracy of the backgrounds is sufficient to guarantee the BRST invariance of the matter action separately, 
as we can see in the following demonstration. First, we act with the BRST operator in the matter part of the action,
\begin{align}\label{prove1}
& Q_1\int d^2\tau\ A_{\mu\nu}\partial_+Z^\mu\partial_-Z^\nu\nonumber\\
& = \int d^2\tau\ \Big( B^{ba}\Lambda_bt^\rho_a\partial_\rho A_{\mu\nu}\partial_+Z^\mu\partial_-Z^\nu + A_{\mu\nu}\partial_+(B^{ba}\Lambda_bt_a^\mu)\partial_-Z^\nu + A_{\mu\nu}\partial_+Z^\mu\partial_-(B^{ba}\Lambda_bt_a^\nu) \Big)\nonumber\\
& = \int d^2\tau\ \Big[ B^{ba}\Lambda_b\Big(t^\rho_a\partial_\rho A_{\mu\nu}\partial_+Z^\mu\partial_-Z^\nu + A_{\mu\nu}\partial_+(t_a^\mu)\partial_-Z^\nu + A_{\mu\nu}\partial_+Z^\mu\partial_-(t_a^\nu) \Big)\\
&\hspace{4cm} - A_{\mu\nu}t^\mu_aB^{ab}\partial_+\Lambda_b\partial_-Z^\nu - A_{\mu\nu}t^\mu_aB^{ab}\partial_+Z^\mu\partial_-\Lambda_b\Big].\nonumber
\end{align}

Next, we use the invariance of the action under the rotations of the 3 planes explained in section \ref{symmetries}. 
This can be implemented by saying that the action is invariant under the action of the $t_a$ generators for $a=12,34,56$,
\begin{equation}
t_a\int d^2\tau\ A_{\mu\nu}\partial_+Z^\mu\partial_-Z^\nu = \int d^2\tau\Big(t_a^\rho\partial_\rho A_{\mu\nu}\partial_+Z^\mu\partial_-Z^\nu + A_{\mu\nu}\partial_+(t_a^\mu)^\mu\partial_-Z^\nu + A_{\mu\nu}\partial_+Z^\mu\partial_-(t_a^\nu) \Big)=0.
\end{equation}
Using this symmetry, equation (\ref{prove1}) reduces to
\begin{equation}
Q_1\int d^2\tau\ A_{\mu\nu}\partial_+Z^\mu\partial_-Z^\nu=\int d^2\tau\ \Big( A_{\mu\nu}t^\mu_aB^{ab}\partial_+\Lambda_b\partial_-Z^\nu + A_{\mu\nu}t^\mu_aB^{ab}\partial_+Z^\mu\partial_-\Lambda_b\Big)\;,
\end{equation}
which is zero from equation (\ref{TA=0}). Therefore, applying $Q_1$ to equation (\ref{Sf}), we find that $Q_1S_g=0$ off-shell. Therefore, since the BRST variation of $S_g$ with respect to $w$ is an expression only involving $\lambda$-ghosts, it cannot cancel with the other terms in $Q_1S_g$, which are expressions containing $w$-ghosts. Therefore, we conclude that
\begin{equation}
Q_w^\alpha\frac{\delta}{\delta w^\alpha}S_g=0, \ \ \ \ \text{off-shell}.
\end{equation}
The only possible way this to be true for every field configuration is if BRST variation of $w$ is zero, that is, $Q_w^\alpha=0$.

This provides a clarification for the BRST variation of the $w$-ghost being zero. Indeed, another way to see this is by looking for 
the structure of the matrix $\kappa$ in the large gamma limit and then seeing what happens to $Q_w$ in (\ref{defQ}) in this limit. 
We can do this by using the coordinates $\{v_0,v_+,v_-\}$ described in section \ref{diagonalization}. In this case, according to 
Equation (\ref{kappaU}), the matrix $\kappa$ in the large $\gamma$ limit becomes:
\begin{equation}\label{kappamatrix}
\kappa \xrightarrow[\gamma\rightarrow\infty]{} (v_-,Bv_+)\begin{pmatrix}
0&0&0\\
0&0&-\frac{1}{\Gamma_+}\\
0&\frac{1}{\Gamma_-}&0\\
\end{pmatrix}\;,
\end{equation}
where $\Gamma_\pm=\alpha\pm\beta$, with $\alpha$ and $\beta$ defined in (\ref{alphabeta}). Therefore, since $Q_w$ 
in (\ref{defQ}) depends only on $\kappa$ acting on $j_a$ and $Ad(t_a)$, and from (\ref{kappamatrix}) we have that 
$\frac{\kappa}{\gamma}\xrightarrow[]{}0$, we conclude that $\frac{Q_w}{\gamma}\xrightarrow[]{}0$. 
Which implies that the BRST variation of $w$ in the limit we are considering is zero.

\section{Conclusion}

In this paper we established a clearer a connection between the string description of the beta deformation and the supergravity solution originally proposed by Maldacena and Lunin in \cite{Lunin_2005}. This was done by first writing the string action with all order deformations and then extracting the supergravity solution.

Formulating the beta deformation in terms of a string action enabled us to investigate the particular problem of strong deformations. Studying the theory in the large $\gamma$ limit was achievable through the presence of a BRST operator defined by $Q_1=B^{ab}\Lambda_at_b$. The final conclusion, represented by the equations in (\ref{deg}), is that the background fields present a degeneracy of dimension 2, given by a particular combinations of the $t_a$ vectors, which rotate the sphere planes in $\mathbb{R}^6$. In terms of the ghost sector, we see that this degeneracy implies that the BRST operator acts trivially in the $w$-ghost. Furthermore, we concluded that as a result of the strong deformation, the geometry of the deformed sphere $\mathbb{S}_{def}^5$ becomes ``saturated" as $\gamma$ approaches infinity, indicating the emergence of a region where the metric degenerates, as indicated in the action (\ref{f123}).

\acknowledgments

This work was supported in part by ICTP-SAIFR FAPESP grant 2019/21281-4 and by FAPESP grant 2022/00940-2. I would like to thank Pedro Vieira for suggesting to study the string description of fishnet limit and for usefull discussions. I would like to thank A. Mikhailov for supervision of this project. I also want to thank Horațiu Năstase and Marcelo Barbosa for the discussions about TsT and Penrose limit, and J. Gomide for english.

\appendix
\section{$AdS_5\times S^5$ embedding coordinates}\label{emb}

The isometry group of the $AdS_5$ is $SU(2,2)$, while the isometry group of the $S^5$ is $SU(4)$. The isometry group of the full super-$AdS_5\times S^5$ is $PSU(2,2|4)$. The algebra of $PSU(2,2|4)$ has a $\mathbb{Z}_4$ gradding: $\mathfrak{g}=\mathfrak{g}_0\oplus\mathfrak{g}_1\oplus\mathfrak{g}_2\oplus\mathfrak{g}_3$. The bosonic part of this algebra is $\mathfrak{g}_0\oplus\mathfrak{g}_2=\mathfrak{p}(\mathfrak{su}(2,2)\oplus\mathfrak{su}(4))$, and $\mathfrak{g}_0=\mathfrak{so}(5)\oplus\mathfrak{so}(1,4)$. The bosonic part of the $PSU(2,2|4)$ group can be written as the following matrix \cite{Metsaev-AdS}:

\begin{equation}
g =\left[
\begin{array}{c;{1pt/1pt}c}
Q & 0   \\ \hdashline[1pt/1pt]
0 & R   \\
\end{array}
\right]
\end{equation}
where $Q\in SU(2,2)$ and $R\in SU(4)$. Let us restrict to the sphere, which means restricting to $SU(4)$. The sphere can be characterized by the coset $S^5=SU(4)/G_0$, with $G_0$ the stabilizer of the 5-sphere given by $SP(4)\cong SO(5)$. Let $E$ be a symplectic structure (that is $E^2=-1$). By definition, $H\in Sp(4)$ when $HEH^T = E$. Using that, we see that $RER^T$ is a good parametrization of $S^5$, since it is invariant under the sphere stabilizer: $R\rightarrow RH$, with $H\in SP(4)$. We then can define the following embedded coordinates:
\begin{equation}
\mathbb{U} := u_j\Sigma_j = RER^T
\end{equation}
with (see \cite{Pedro} and the references therein):
\begin{equation}\label{uS}
u_j\Sigma_j = \left(
\begin{array}{cccc}
 0 & -u_6-i u_5 & -u_4-i u_3 & -u_2-i u_1 \\
 u_6+i u_5 & 0 & -u_2+i u_1 & u_4-i u_3 \\
 u_4+i u_3 & u_2-i u_1 & 0 & -u_6+i u_5 \\
 u_2+i u_1 & -u_4+i u_3 & u_6-i u_5 & 0 \\
\end{array}
\right)
\end{equation}
where $\Sigma_j$ are the gamma matrices for $SO(6)$, which translates representation \textbf{6} into $\textbf{4}\wedge\textbf{4}$. The generators of the symmetry algebra of the sphere $\mathfrak{su}(4)$ in the embedding coordinates in $\textbf{6}\wedge\textbf{6}$ are:
\begin{equation}\label{tij}
t_{ij} = u_{[i}\frac{\partial}{\partial u^{j]}}
\end{equation}
and the structure constants are given by:
\begin{equation}
f_{ij,kl}^{\ \ \ \ mn} = \delta_{n[l}\delta_{k]
[j}\delta_{i]m}-\delta_{m[l}\delta_{k]
[j}\delta_{i]n} - \delta_{n[j}\delta_{i][l}\delta_{k]m}+\delta_{m[j}\delta_{i][l}\delta_{k]n}
\end{equation}
We may also translate it to:
\begin{equation}
t_{ij}\longmapsto \mathbb{T}_{ij}=\Sigma_i\cdot\Sigma_j
\end{equation}
the generators of $\mathfrak{su}(4)$ will be:
\begin{align}\label{generators}
&\mathbb{T}_{12} = \left(
\begin{array}{cccc}
 -i & 0 & 0 & 0 \\
 0 & i & 0 & 0 \\
 0 & 0 & i & 0 \\
 0 & 0 & 0 & -i \\
\end{array}
\right),\ \mathbb{T}_{13} = \left(
\begin{array}{cccc}
 0 & 1 & 0 & 0 \\
 -1 & 0 & 0 & 0 \\
 0 & 0 & 0 & -1 \\
 0 & 0 & 1 & 0 \\
\end{array}
\right),\ \mathbb{T}_{14} = \left(
\begin{array}{cccc}
 0 & i & 0 & 0 \\
 i & 0 & 0 & 0 \\
 0 & 0 & 0 & -i \\
 0 & 0 & -i & 0 \\
\end{array}
\right)\nonumber\\
&\mathbb{T}_{15} = \left(
\begin{array}{cccc}
 0 & 0 & -1 & 0 \\
 0 & 0 & 0 & -1 \\
 1 & 0 & 0 & 0 \\
 0 & 1 & 0 & 0 \\
\end{array}
\right),\ \mathbb{T}_{16} = \left(
\begin{array}{cccc}
 0 & 0 & -i & 0 \\
 0 & 0 & 0 & -i \\
 -i & 0 & 0 & 0 \\
 0 & -i & 0 & 0 \\
\end{array}
\right),\   \mathbb{T}_{23} = \left(
\begin{array}{cccc}
 0 & -i & 0 & 0 \\
 -i & 0 & 0 & 0 \\
 0 & 0 & 0 & -i \\
 0 & 0 & -i & 0 \\
\end{array}
\right),\nonumber\\
&\mathbb{T}_{24}=\left(
\begin{array}{cccc}
 0 & 1 & 0 & 0 \\
 -1 & 0 & 0 & 0 \\
 0 & 0 & 0 & 1 \\
 0 & 0 & -1 & 0 \\
\end{array}
\right),\ \mathbb{T}_{25} = \left(
\begin{array}{cccc}
 0 & 0 & i & 0 \\
 0 & 0 & 0 & -i \\
 i & 0 & 0 & 0 \\
 0 & -i & 0 & 0 \\
\end{array}
\right),\ \mathbb{T}_{26} = \left(
\begin{array}{cccc}
 0 & 0 & -1 & 0 \\
 0 & 0 & 0 & 1 \\
 1 & 0 & 0 & 0 \\
 0 & -1 & 0 & 0 \\
\end{array}
\right),\\
&\mathbb{T}_{34} = \left(
\begin{array}{cccc}
 -i & 0 & 0 & 0 \\
 0 & i & 0 & 0 \\
 0 & 0 & -i & 0 \\
 0 & 0 & 0 & i \\
\end{array}
\right),\ \mathbb{T}_{35} = \left(
\begin{array}{cccc}
 0 & 0 & 0 & 1 \\
 0 & 0 & -1 & 0 \\
 0 & 1 & 0 & 0 \\
 -1 & 0 & 0 & 0 \\
\end{array}
\right),\ \mathbb{T}_{36} = \left(
\begin{array}{cccc}
 0 & 0 & 0 & i \\
 0 & 0 & -i & 0 \\
 0 & -i & 0 & 0 \\
 i & 0 & 0 & 0 \\
\end{array}
\right),\nonumber\\
&\mathbb{T}_{45} = \left(
\begin{array}{cccc}
 0 & 0 & 0 & -i \\
 0 & 0 & -i & 0 \\
 0 & -i & 0 & 0 \\
 -i & 0 & 0 & 0 \\
\end{array}
\right),\ \mathbb{T}_{46} = \left(
\begin{array}{cccc}
 0 & 0 & 0 & 1 \\
 0 & 0 & 1 & 0 \\
 0 & -1 & 0 & 0 \\
 -1 & 0 & 0 & 0 \\
\end{array}
\right),\ \mathbb{T}_{56} = \left(
\begin{array}{cccc}
 -i & 0 & 0 & 0 \\
 0 & -i & 0 & 0 \\
 0 & 0 & i & 0 \\
 0 & 0 & 0 & i \\
\end{array}
\right)\nonumber
\end{align}

Elements in $G_0$ are characterized by $H = EH^TE$. Therefore, we can write a generic even element $M$ as a sum $M=M_0+M_2$ with $M_0 = EM^TE$ and $M_2 = M-EM^TE$. For example, the current of the sigma model is given by $R^TdR$ and we may project it:
\begin{equation}
(R^{T}dR)_0 = -E(R^TdR)E\ ,\ \ \ \ \ (R^{T}dR)_2 = R^{T}dR + E(R^TdR)E
\end{equation}
It is remarkable that the projection of the current in $\Bar{2}$ can be related to an expression with $\mathbb{U}$:
\begin{equation}
\mathbb{U}^Td\mathbb{U} = -( RER^TdR ER^T + dRR^T ) = - R(R^TdR)_2R^T
\end{equation}

The bosonic part of the Lagrangian for the sigma-model is $\Str[J_2\wedge\star J_2]$, where $J_2$ is the $\Bar{2}$-component of the current $J = -g^{-1}dg$. The subsector of the sphere $S^5$ will have the current given by $J = -R^TdR$. Therefore we set an important relation:
\begin{equation}
\mathbb{U}^Td\mathbb{U} = RJ_2R^T = j_2
\end{equation}
which is the bosonic part of the density of current. When we compute it we find:
\begin{equation}\label{j2}
j_2 = \sum_{0\le i<j\le6} u_{[i}du_{j]}\mathbb{T}_{ij}
\end{equation}
Using this, we see that the action is given by the product of two $\mathbb{U}^Td\mathbb{U}$:
\begin{equation}\label{action}
S = \tr[\mathbb{U}^Td\mathbb{U} \wedge \star \mathbb{U}^Td\mathbb{U}] = \tr [RJ_2R^T\wedge\star RJ_2R^T] = \tr [J_2\wedge\star J_2]
\end{equation}\\
In the same way we define the action of the lie algebra in $R$ by $\delta_aR=t_aR$, where $t_a$ are algebra generators. And then follows:
\begin{equation}
\mathbb{U}^T\delta_a\mathbb{U} = -R(R^T\delta_aR)_2R^T
\end{equation}
and therefore:
\begin{equation}\label{udu}
\Tr\Big[(\mathbb{U}^T\delta_a\mathbb{U})(\mathbb{U}^T\delta_b\mathbb{U})\Big] = \Tr\Big[ (R^T\delta_aR)_2(R^T\delta_bR)_2 \Big] =: M^{(2)}_{ab}
\end{equation}
We use this to compute the bosonic part of $M$ explicitly. Let us compute it when $a$ are indices of the $\mathfrak{su}(4)$ lie algebra. We use the matrix $\mathbb{U}$ in (\ref{uS}) and the generators $t_{ij}$ in the form of a vector field, as written in (\ref{tij}). Applying this to the left hand side of (\ref{udu}) we find the following non-vanishing components for $M^{(2)}$:
\begin{equation}\label{mij}
M^{(2)}_{[ij][ij]} = (u_i^2+u_j^2)
\end{equation}
\begin{equation}\label{mij2}
M^{(2)}_{[ij][ik]} = (u_ju_k),\ \ \text{for}\ j\ne k
\end{equation}

\subsection{Grading of the algebra}\label{B1}
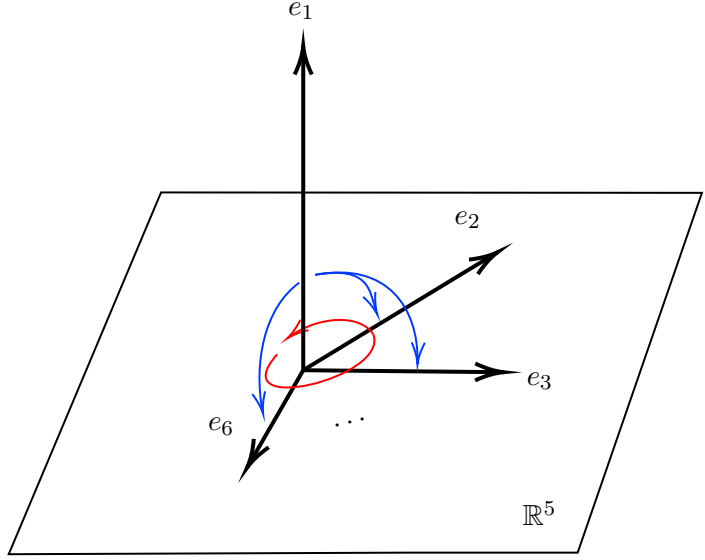
\begin{figure}
    \centering
\tikzset{every picture/.style={line width=0.75pt}} %set default line width to 0.75pt        

\begin{tikzpicture}[x=0.75pt,y=0.75pt,yscale=-1,xscale=1]
%uncomment if require: \path (0,495); %set diagram left start at 0, and has height of 495

%Straight Lines [id:da16422512167820202] 
\draw [line width=1.5]    (275,233) -- (369.5,175.12) ;
\draw [shift={(372.06,173.56)}, rotate = 148.51] [color={rgb, 255:red, 0; green, 0; blue, 0 }  ][line width=1.5]    (14.21,-4.28) .. controls (9.04,-1.82) and (4.3,-0.39) .. (0,0) .. controls (4.3,0.39) and (9.04,1.82) .. (14.21,4.28)   ;
%Straight Lines [id:da8359334491143467] 
\draw [line width=1.5]    (275,233) -- (248.49,279.4) ;
\draw [shift={(247,282)}, rotate = 299.74] [color={rgb, 255:red, 0; green, 0; blue, 0 }  ][line width=1.5]    (14.21,-4.28) .. controls (9.04,-1.82) and (4.3,-0.39) .. (0,0) .. controls (4.3,0.39) and (9.04,1.82) .. (14.21,4.28)   ;
%Straight Lines [id:da46719066802629483] 
\draw [line width=1.5]    (275,233) -- (372,233.97) ;
\draw [shift={(375,234)}, rotate = 180.57] [color={rgb, 255:red, 0; green, 0; blue, 0 }  ][line width=1.5]    (14.21,-4.28) .. controls (9.04,-1.82) and (4.3,-0.39) .. (0,0) .. controls (4.3,0.39) and (9.04,1.82) .. (14.21,4.28)   ;
%Curve Lines [id:da6808957121938488] 
\draw [color={rgb, 255:red, 255; green, 0; blue, 0 }  ,draw opacity=1 ]   (262,225) .. controls (234,257) and (318,237) .. (310,215) ;
%Curve Lines [id:da10041143214103199] 
\draw [color={rgb, 255:red, 255; green, 0; blue, 0 }  ,draw opacity=1 ]   (310,215) .. controls (303.25,204.39) and (282.52,205.88) .. (268.5,215.88) ;
\draw [shift={(267,217)}, rotate = 321.84] [color={rgb, 255:red, 255; green, 0; blue, 0 }  ,draw opacity=1 ][line width=0.75]    (10.93,-3.29) .. controls (6.95,-1.4) and (3.31,-0.3) .. (0,0) .. controls (3.31,0.3) and (6.95,1.4) .. (10.93,3.29)   ;
%Straight Lines [id:da9152909928697132] 
\draw [line width=1.5]    (275,233) -- (275,73) ;
\draw [shift={(275,70)}, rotate = 90] [color={rgb, 255:red, 0; green, 0; blue, 0 }  ][line width=1.5]    (14.21,-4.28) .. controls (9.04,-1.82) and (4.3,-0.39) .. (0,0) .. controls (4.3,0.39) and (9.04,1.82) .. (14.21,4.28)   ;
%Shape: Polygon [id:ds33882778050531204] 
\draw   (204.06,143.56) -- (473.93,143.76) -- (411.93,324.76) -- (327.93,324.76) -- (128.06,325.56) -- cycle ;
%Curve Lines [id:da6946975531344366] 
\draw [color={rgb, 255:red, 0; green, 58; blue, 255 }  ,draw opacity=1 ]   (281,185) .. controls (317.26,178.14) and (332.39,202.01) .. (332.04,228.38) ;
\draw [shift={(332,230)}, rotate = 272.12] [color={rgb, 255:red, 0; green, 58; blue, 255 }  ,draw opacity=1 ][line width=0.75]    (10.93,-3.29) .. controls (6.95,-1.4) and (3.31,-0.3) .. (0,0) .. controls (3.31,0.3) and (6.95,1.4) .. (10.93,3.29)   ;
%Curve Lines [id:da031314362434450516] 
\draw [color={rgb, 255:red, 0; green, 58; blue, 255 }  ,draw opacity=1 ]   (281,185) .. controls (308.26,180.3) and (308.13,193.27) .. (311.34,203.15) ;
\draw [shift={(312,205)}, rotate = 248.2] [color={rgb, 255:red, 0; green, 58; blue, 255 }  ,draw opacity=1 ][line width=0.75]    (10.93,-3.29) .. controls (6.95,-1.4) and (3.31,-0.3) .. (0,0) .. controls (3.31,0.3) and (6.95,1.4) .. (10.93,3.29)   ;
%Curve Lines [id:da6633467714826371] 
\draw [color={rgb, 255:red, 0; green, 58; blue, 255 }  ,draw opacity=1 ]   (273,189) .. controls (255.45,201.68) and (250.26,229.56) .. (254.64,253.19) ;
\draw [shift={(255,255)}, rotate = 258.23] [color={rgb, 255:red, 0; green, 58; blue, 255 }  ,draw opacity=1 ][line width=0.75]    (10.93,-3.29) .. controls (6.95,-1.4) and (3.31,-0.3) .. (0,0) .. controls (3.31,0.3) and (6.95,1.4) .. (10.93,3.29)   ;

% Text Node
\draw (266,46.4) node [anchor=north west][inner sep=0.75pt]    {$e_{1}$};
% Text Node
\draw (349,151.4) node [anchor=north west][inner sep=0.75pt]    {$e_{2}$};
% Text Node
\draw (385,232.4) node [anchor=north west][inner sep=0.75pt]    {$e_{3}$};
% Text Node
\draw (287.56,257.26) node [anchor=north west][inner sep=0.75pt]  [rotate=-345.56]  {$\cdots $};
% Text Node
\draw (226,255.4) node [anchor=north west][inner sep=0.75pt]    {$e_{6}$};
% Text Node
\draw (383,297.4) node [anchor=north west][inner sep=0.75pt]    {$\mathbb{R}^{5}$};

\end{tikzpicture}
    \caption{Representation of $SO(6)$ rotations using embedding coordinates of $\mathbb{R}^6$. In red we have the $\mathfrak{so}(6)_0$ part while in blue the $\mathfrak{so}(6)_2$ part.}
    \label{axis}
\end{figure}

We have described the isometry of the sphere as the $\mathfrak{su}(6)$ group with generators given by $\{t_{ij}\}_{1\le i<j\le6}$. This is a subalgebra of the $\mathfrak{psu}(2,2|4)$. This algebra splits into $\mathfrak{g}_0\oplus\mathfrak{g}_1\oplus\mathfrak{g}_2\oplus\mathfrak{g}_3$, where $\mathfrak{g}_0$ is the subalgebra of stabilizer of $AdS_5\times S^5$. We want to write $\mathfrak{so}(4)_0\subset\mathfrak{g}_0$ in terms of the $t_{ij}$ elements. To do that, we first choose a direction in the $\mathbb{R}^6$, for instance $e_1$. Then, the part of the algebra inside $\mathfrak{g}_0$ will be the rotations that let $e_1$ invariant:
\begin{equation}
\mathfrak{su}(4)_0 = \{ t_{ij}\ |\ i,j\ne1 \}
\end{equation}
and therefore, the $\mathfrak{su}(4)_2$ is generated by the complementary algebra:
\begin{equation}
\mathfrak{su}(4)_2 = \big\{ t_{12},t_{13}, t_{14}, t_{15}, t_{16}\big\}
\end{equation}
we can check the grading: $[t_{1i},t_{ij}]=t_{1j}$, $[t_{ij},t_{kl}]=f_{ij,kl}^{mn}t_{mn}$ and $[t_{ij},t_{1k}]=\delta_{jk}t_{1i}-\delta_{ik}t_{1j}$. This grading is pictorially represented in Figure \ref{axis}. We than rewrite the indices of $\mathfrak{su}(4)_2$ as:
\begin{equation}
t_{1j} \rightarrow t_j
\end{equation}

Based on that we can prove that the map $\Ad_g$ (\ref{surjmap}) is surjective. To do that, we first restrict to the left sector, that is, $u=u_L$ (i.e., equation (\ref{AdL})). Then, the same thing can be done for the right sector, $u=u_R$ (i.e., equation (\ref{AdR})). We also use the corresponding infinitesimal map $\ad_Z$:
\begin{align}
& \ad_Z: u \mapsto \sum_{a=12,34} [u,Z]^a t_a=  u^\alpha Z^{\beta} f_{\alpha\beta}^{2}t_{2} + u^\alpha Z^{\hat\alpha} f_{\alpha\hat\alpha}^{56}t_{56}
\end{align}
We can prove it for the infinitesimal map and then integrate in the whole group using the exponential map: $\Ad_g=\Ad_{\exp[Z]}=\exp[\ad_Z]$. We can to this because $SU(6)$ is compact and then the exponential map will be surjective. The structure constants are:
\begin{equation}
f_{\alpha\beta}^{j} = \gamma_{\alpha\beta}^{j}\ \ \ \ f_{\alpha\hat\alpha}^{ij} = \gamma_{\alpha\hat\alpha}^{ij}=[\gamma^i,\gamma^j]_{\alpha\hat\alpha}
\end{equation}
with $\gamma^j$ the gamma matrices in $10D$. Take a vector $Z^\mu$. We want to pick two different spinors and send it to two different 2-dimensional vectors generated by $t_{12}$ and $t_{34}$. First, take a spinor $u$ such that:
\begin{equation}
u_\alpha Z^\alpha=0
\end{equation}
since $\gamma^2$ is invertible, the following subspace has co-dimension one:
\begin{equation}
S = \Bigg\{ u^\alpha,\ \ u^\alpha\gamma^2_{\alpha\beta}Z^\beta=0 \Bigg\}
\end{equation}
now, we send this subspace to:
\begin{equation}
\hat S = \Bigg\{ u_{\hat\alpha}=u^\alpha\gamma^{34}_{\alpha\hat\alpha},\ \ \ u^\alpha\in S \Bigg\}
\end{equation}
which has also co-dimension one since $\gamma^{34}$ is also invertible. We therefore can adjust $Z$ such that:
\begin{equation}\label{adj}
\bar u_{\hat\alpha}Z^{\hat\alpha}=\bar u^\alpha\gamma^{34}_{\alpha\hat\alpha}Z^{\hat\alpha}\ne0,
\end{equation}
for some $\bar u\in\hat{S}$. Therefore, the action of $\bar u$ under our map covers only the space generated by $t_{34}$. In a similar way we can find another $\Bar{\bar u}$ such that the action under our 
map only covers $t_{12}$. Now, if we want to apply this idea to (\ref{isometry1}), we need to consider arbitrary $Z^\mu$. Here, the initial $Z^\mu$ was arbitrary but we had to modify the $Z^{\hat\alpha}$ part of it in order to satisfy equation (\ref{adj}) since it is not necessarily satisfied. This adjustment happens when the $Z^{\hat\alpha}$ part belongs to $\hat{S}^\perp$, that is, when $u^\alpha\gamma^{34}_{\alpha\hat\alpha}Z^{\hat\alpha}=0$ for every $u\in S$. One possible adjustment is done by taking $Z_0^{\hat\alpha}\in\hat{S}$ and modify $Z^{\hat\alpha}$ by:
\begin{equation}
Z^{\hat\alpha}(\epsilon)=Z^{\hat\alpha} +\epsilon Z_0^{\hat\alpha},
\end{equation}
where $\epsilon$ is a non-zero free parameter. We therefore obtain equation (\ref{isometry}) at the space-time point $ Z^\mu(\epsilon) = Z^\mu +\epsilon Z_0^{\hat\alpha}$, where $Z^\mu$ is an arbitrary point in space-time but $Z_0^{\hat\alpha}$ is not. Since we will have:
\begin{equation}
(A_{\mu\nu}t_a^\mu)\Big|_{Z=Z(\epsilon)}=0\ \ \forall \epsilon\ne0
\end{equation}
we take the limit $\epsilon\rightarrow0$ and obtain the equation valid for all points in space-time.

In order to go from the infinitesimal case to what we really have, we use the relation:
\begin{equation}
g^{-1}ug = u^\alpha W_\alpha^{\ A}t_A = u^\alpha Z^B w_\alpha^C f^A_{BC}t_A,
\end{equation}
where $W = \frac{1-e^{\ad_X}}{\ad_X}$, and we used that the structure constants we are using are invertible matrices to rewrite $W$ in terms of a new $w$ and the coordinates $Z$. Then, using what we already found for the infinitesimal case, we find the space of $u$'s such that $u^\alpha W_\alpha^{12}=0$ and inside this space we choose a specific one such that $\bar u^\alpha W_\alpha^{34}\ne0$, and vice-versa.

\section{Penrose Limit}\label{penrose}

Consider the following null geodesic:

\begin{align}
&u_5+iu_6 =  e^{i\mathcal{J}\tau} = \cos(\mathcal{J}\tau) + i\sin (\mathcal{J}\tau):\ \ \ \ \ \tau\in\mathbb{R}\longmapsto S^5\\
&v_5+iv_6 =  e^{i\mathcal{E}\tau} = \cos(\mathcal{E}\tau) + i\sin (\mathcal{E}\tau):\ \ \ \ \ \tau\in\mathbb{R}\longmapsto AdS_5
\end{align}

Let us focus attention on the sphere part. If we parameterize the other sphere coordinates by $u_1+iu_2=\nu e^{i\theta}$ and $u_3+iu_4=\rho e^{i\psi}$ and use equation (\ref{action}) to find the action near the geodesic:

\begin{equation}
S = R^2\int_{\Sigma}d^2\sigma  \left(1+\rho^2\right)(\partial\nu)^2+  \left(1+\nu^2\right)(\partial\rho)^2-\nu\rho (\partial\nu\bar\partial\rho+\partial\rho\bar\partial\nu)\nonumber
\end{equation}
\begin{equation}
+ \left(\nu^2+\rho^2+1\right) \left( J^2 +
\nu^2(\partial\theta)^2+ \rho^2(\partial\psi)^2\right) + [AdS_5]
\end{equation}

and now we parameterize $\nu\rightarrow\nu/R$ and
$\rho\rightarrow\rho/R$ and take the limit $R\rightarrow\infty$ to obtain:
\begin{equation}
S_{S^5} = \int_{\Sigma}d^2\sigma\Big( (\partial\nu)^2 + \nu^2(\partial\theta)^2 + (\partial\rho)^2 + \rho^2(\partial\psi)^2  +  J^2\left(\nu^2+\rho^2\right)\Big)
\end{equation}
\begin{equation}
= \int_{\Sigma}d^2\sigma\left( \sum_{i=1}^4(\partial u_i)^2 + J^2\left(\sum_{i=1}^4 u_i^2\right)\right)
\end{equation}

% Bibliography

%% [A] Recommended: using JHEP.bst file
%\bibliographystyle{JHEP.bst}
%\bibliography{biblio.bib}

\providecommand{\href}[2]{#2}\begingroup\raggedright\endgroup

%% or
%% [B] Manual formatting (see below)
%% (i) We suggest to always provide author, title and journal data or doi:
%% in short all the informations that clearly identify a document.
%% (ii) please avoid comments such as "For a review'', "For some examples",
%% "and references therein" or move them in the text. In general, please leave only references in the bibliography and move all
%% accessory text in footnotes.
%% (iii) Also, please have only one work for each \bibitem.

\end{document}